\documentclass[
	final,
	aps,
	pre,
	twocolumn,
	]{revtex4-1}
\let\Twocolumn

\newif\ifTwocolumn
\ifx\Twocolumn\undefined
  \Twocolumnfalse
\else
  \Twocolumntrue
\fi
\ifTwocolumn
\fi
\usepackage{graphicx}
\usepackage{xcolor}
\usepackage{amsmath,amssymb,amsfonts}
\usepackage{gensymb}
\usepackage{bm}
\usepackage{upgreek}
\usepackage{wasysym}
\usepackage{bibentry}
\usepackage[T1]{fontenc}
\usepackage{textcomp}
\usepackage{mathptmx}
\usepackage[scaled=0.92]{helvet}
\usepackage{courier}
\usepackage{ulem}
\newcommand{\eref}[1]{Eq.~\eqref{#1}}
\newcommand{\fref}[1]{Fig.~\ref{#1}}

\newcommand{\upd}{{\ensuremath{\textrm{d}}}}
\renewcommand*{\vec}[1]{\mathbf{#1}}

\newcommand{\Pp}{\ensuremath{(+,+)}}

\newcommand{\Dpp}{\ensuremath{\Delta_{\Pp}}}

\newcommand{\temp}{\ensuremath{t}}

\ifTwocolumn
  
\else
  
\fi

\usepackage{setspace}
\ifTwocolumn
\newcommand{\clevercaption}[1]{\caption{\setstretch{1.0}#1}}
\else
\newcommand{\clevercaption}[1]{\caption{\setstretch{1.2}#1}}
\fi

\begin{document}
\title{Critical Casimir forces between planar and crenellated surfaces}
\author{M.~Tr{\"o}ndle}
\email{troendle@is.mpg.de}
\author{L.~Harnau}
\email{harnau@is.mpg.de}
\author{S.~Dietrich}
\email{dietrich@is.mpg.de}
\affiliation{
	Max-Planck-Institut f\"ur Intelligente Systeme,  
	Heisenbergstr.\ 3, D-70569 Stuttgart, Germany}
\affiliation{
	Institut f\"ur Theoretische Physik IV, 
	Universit\"at Stuttgart, 
	Pfaffenwaldring 57, 
	D-70569 Stuttgart, Germany
}
\date{\today}
%
\begin{abstract}
  We study critical Casimir forces between planar walls and geometrically structured substrates
  within mean-field theory. As substrate structures, crenellated surfaces consisting of periodic 
arrays of rectangular crenels and merlons are considered. Within the widely used proximity force 
approximation, both the top surfaces of the merlons and the bottom surfaces of the crenels contribute 
to the critical Casimir force. However, for such systems the full, numerically determined critical Casimir forces deviate 
significantly from the pairwise addition formalism underlying the proximity force approximation. 
A first-order correction to the proximity force approximation is presented in terms of a step 
contribution arising from the critical Casimir interaction between a planar substrate and the 
right-angled steps of the merlons consisting of their upper and lower edges as well as their sidewalls. 
\end{abstract}
\maketitle
\section{Introduction \label{sec:intro}}
The thermodynamic analogue of the Casimir effect originating from the confinement of vacuum
fluctuations \cite{Casimir:1948,Capasso:2007} is the critical Casimir effect due to the presence
of long-ranged thermal fluctuations in a fluid close to its critical point at ${T=T_c}$.
The corresponding critical Casimir forces have been predicted theoretically by Fisher and de Gennes in 1978 
\cite{Fisher:1978,Krech:book,Brankov:book,Gambassi:2009conf}.
Experimentally, critical Casimir forces have been studied only during the last decade when
first measurements were performed indirectly via monitoring the thickness of
wetting films upon approaching a critical end point \cite{Garcia:1999,Garcia:2002,Fukuto:2005,Ganshin:2006,Rafai:2007}.
Later on, the critical Casimir force has been measured directly by using colloidal particles suspended in a 
binary liquid mixture \cite{Hertlein:2008,Gambassi:2009,Nellen:2009}.
\par
Generically, the surfaces, which confine a binary liquid mixture, preferentially attract one of its two components 
leading to either positive [$(+)$] or negative [$(-)$] values of the scalar order parameter $\phi$ 
which describes the difference between the local concentration of one of the two components and its 
critical value.
This generic preference of the surfaces confining the liquid can be described by effective surface fields.
Upon approaching $T_c$, the critical adsorption profiles, which describe the concentration enhancement near the surface, 
become long ranged due to the concomitant divergence of the bulk correlation length \cite{Fisher:1978,Binder:1983,Diehl:1986}.
In semi-infinite systems, the transition from the phase in which only the region near the single surface is ordered 
to the one in which also the bulk is ordered is known as the extraordinary or normal transition \cite{bray:1977,burkhardt:1994}.
For two surfaces opposing each other, depending on the mutual combinations of boundary conditions (BCs) critical Casimir forces 
are either attractive or repulsive. Their range is set by the bulk correlation length $\xi$.
Whereas in fluids $\xi$ is typically of molecular size, it attains values of the order of micrometers upon 
approaching the critical point \cite{Hertlein:2008,Gambassi:2009,Nellen:2009}.
Thus, critical fluctuations may induce effective interaction potentials with a strength of several $k_BT$ at the nanometer and micrometer scale.
Moreover, critical Casimir forces are universal in character: due to the divergence of the correlation
length molecular details of the confined binary liquid mixture become irrelevant and only a few gross features of the system
determine the main characteristics of the critical Casimir forces \cite{Krech:book,Brankov:book,Gambassi:2009conf}.
\par
\begin{figure} 
  \ifTwocolumn
  \includegraphics[width=8.0cm,clip=true]{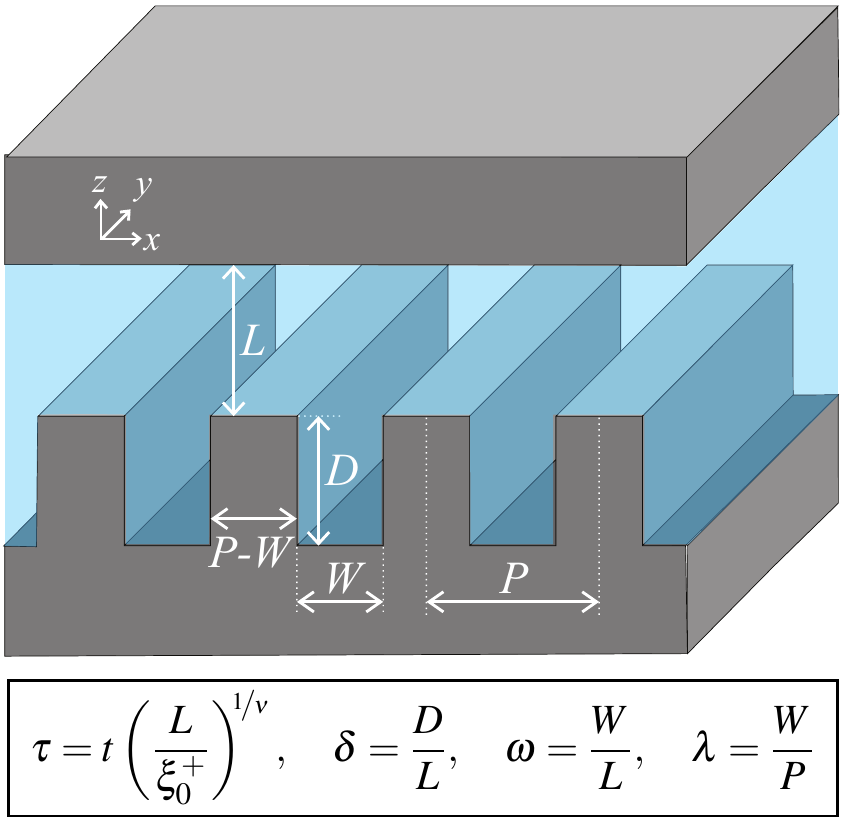} 
  \else
  \includegraphics[width=8.50cm,clip=true]{sketch_final} 
  \fi
  \clevercaption{%
  Sketch of the geometry under consideration. An upper, planar wall is located at a minimal surface-to-surface distance 
  $L$ from a lower, parallel, crenellated wall which exhibits a periodic pattern with 
  period $P$ consisting of rectangular crenels of width $W$ and depth $D$ and merlons of width $P-W$ and 
  height $D$. The system is spatially invariant 
  in the $y$-direction. For later reference, the box at the bottom summarizes the definitions of the 
  various scaling variables which the scaling function of the critical Casimir force depends on;
  $t$ is the reduced temperature $t=(T-T_c)/T_c$ and $\xi_0^+$ is the nonuniversal amplitude of the bulk correlation length $\xi_+(t\to0)=\xi_0^+t^{-\nu}$ in the disordered phase, which defines the universal critical exponent $\nu$.
  }   
  \label{fig:sketch}
\end{figure}
In view of nano- and micro-electromechanical devices, nowadays various experimental techniques are available to endow solid 
surfaces with precisely defined geometrical or chemical structures 
(see, e.g., Refs.~~\cite{Xia:1998,Wang:1998,Tolfree:1998,Nie:2008,Thorsen:2002}).
Critical Casimir forces for \emph{chemically} structured confinements have been studied theoretically 
\cite{Sprenger:2006,Troendle:2009,Troendle:2010,Gambassi:2011,Parisen:2010,Parisen:2013,Parisen:2014,Labbe:2014} as well as
in experiments with colloidal particles \cite{Soyka:2008,Troendle:2011}.
It has been demonstrated that such patterns induce \emph{lateral} critical Casimir forces which can be used to trap particles reversibly 
along the lateral direction in a designated way \cite{Soyka:2008,Troendle:2011}.
Moreover, a suitable combination of attractive and repulsive critical Casimir forces may even lead
to stable levitation \cite{Troendle:2010,Gambassi:2011}.
\par
Here, we study critical Casimir forces for \emph{geometrically} structured confinements.
The quantum-electrodynamic Casimir effect in the presence of geometrically structured surfaces has been 
studied theoretically and experimentally
for various surface topographies \cite{Emig:2001,Emig:2003,Zwol:2008,Lambrecht:2008,Rodriguez:2011,Lussange:2012}.
Two opposing surface gratings are subject to lateral quantum electrodynamic Casimir forces \cite{Chen:2002,Emig:2003,
Emig:2005,Rodrigues:2006,Dalvit:2008,Ashourvan:2007,Emig:2007,Miri:2008,Rodrigues:2008,Chiu:2010}.
In this context, experimental studies focus on spherical particles near crenellated surfaces, i.e., forming 
grooves with rectangular cross-sections \cite{Chan:2008,Bao:2010,Guerout:2013,Intravaia:2013}. Typically, the radius of the spherical particles 
is much larger than the period of the pattern of the crenellated surface such that, effectively, in the region of 
closest approach the system mimics the geometrical setup of a planar wall near a parallel, crenellated surface.
Critical adsorption and critical Casimir forces for geometrically structured confinements 
have been studied for structures shaped as wedges and ridges with triangular cross-section
\cite{Cardy:1983,Hanke:1999,Palagyi:2004,Troendle:2008}, as well as for truncated wedges \cite{Bimonte:2014}.
It was found that, for large distances between a sawtooth-shaped wall and a planar wall, the
critical Casimir force effectively corresponds to the one between two planar walls; on the other hand,
for short distances between the two surfaces, the tips of the ridges dominate the order parameter
profile and the characteristic power law behavior of the critical Casimir force differs from that for planar surfaces \cite{Troendle:2008}.
\par

The present study extends these previous investigations \cite{Cardy:1983,Hanke:1999,Palagyi:2004,Troendle:2008} into various directions.
We consider a crenellated substrate close to a planar substrate at minimal surface-to-surface distance $L$,
as shown in \fref{fig:sketch}.
The details of the geometry as well as the finite-size scaling of the critical Casimir phenomena
are described in Section~\ref{sec:theory}.
We calculate universal scaling functions for the critical Casimir forces and for the order parameter profiles
within mean-field theory and for identical chemical BCs at both walls 
($(+,+)$ configuration).
In Section~\ref{sec:profiles} we first study the universal features of the order parameter profiles 
close to $T_c$ for the geometry under consideration.
Second, in Section~\ref{sec:force} we study the critical Casimir forces acting on such
geometrically structured substrates.
Finally, in Section~\ref{sec:summary} we summarize our main findings.
\section{Finite-size scaling and Mean-field theory \label{sec:theory}}
%
According to the theory of finite-size scaling, the singular contribution to the critical Casimir force is described by 
a universal scaling function, which is independent of the molecular details of the binary liquid mixture
and depends only on the bulk universality  class of the associated critical point \cite{barber:1983,privman:1990,Krech:book,Brankov:book}.
Here, we focus on the Ising universality class characterized by a scalar order parameter $\phi$, which encompasses
the experimentally relevant binary liquid mixtures and simple fluids.
Upon approaching the critical point of the fluid, the bulk correlation length diverges as
$\xi_\pm(t\to0)=\xi_0^{\pm}|t|^{-\nu}$, where $\nu\simeq0.63$ in spatial dimension $d=3$ and $\nu=1/2$ in 
spatial dimension $d=4$ \cite{pelissetto:2002}; $\xi_0^\pm$ are nonuniversal amplitudes characterized by the 
universal ratio 
$\xi_0^+ / \xi_0^- \simeq 1.9$ in $d=3$ and $\xi_0^+ / \xi_0^- = \sqrt{2}$ in $d=4$. The sign of the 
reduced temperature $t=\pm (T-T_c)/T_c$ is chosen such that $t>0$ corresponds to the mixed (disordered) phase 
of the fluid, whereas $t<0$ corresponds to the ordered phase, corresponding to spontaneous phase separation. 
For an upper critical point the homogeneous phase is found at high temperatures, and one has $t=(T-T_c)/T_c$. 
However, many experimentally relevant binary liquid mixtures exhibit a lower critical point; in this 
case $t=-(T-T_c)/T_c$.
\par
In general, the sign and the amplitude of the critical Casimir force depend on the types of effective chemical BCs 
at the walls and on the geometry of the confining surfaces.
Here we focus on the case of equal, symmetry-breaking $(+)$ BCs, which corresponds
to the generic case of preferential adsorption of one of the two species of a binary liquid mixture. 
This leads to an attractive critical Casimir force \cite{Krech:1997}.
Inspired by experiments encompassing binary liquid mixtures of water and lutidine (with a lower critical point
at $T_c\simeq34\degree C$) \cite{Hertlein:2008,Soyka:2008,Gambassi:2009,Troendle:2011}, here a binary 
liquid mixture  with a lower critical point is considered at fixed pressure and at its critical composition.
\subsection{Planar walls \label{sec:planar}}
%
First, we briefly review the film geometry. In this case the liquid 
is confined between two parallel, macroscopically extended walls at a distance $l$.
According to renormalization group theory the critical Casimir force $f_{\parallel}$
per area of one wall, which is acting on the parallel walls ($\parallel$), scales as \cite{Krech:1991,Krech:1992a,Krech:1992}
\begin{equation} 
  \label{eq:planar-force}
  f_{\parallel}(l,T)=k_BT \frac{1}{l^d}k_{\parallel}( t(l/\xi_0^+)^{1/\nu}).
\end{equation} 
The scaling function $k_{\parallel}$ depends only on a single scaling variable given by
the film thickness $l$ in units of $\xi_\pm$, raised to the power $1/\nu$.
For equal chemical BCs, as discussed here, $k_\parallel$ is negative, so that the critical
Casimir force is attractive.
For $T\to T_c$ the scaling function of the critical Casimir force reduces to a universal 
constant value, the so-called critical Casimir amplitude \cite{Krech:book,Brankov:book}:
\begin{equation} 
  \label{eq:delta-ab}
  k_{\parallel}(0)=\Delta_{(+,+)}.
\end{equation} 
Accordingly, at $T_c$ the critical Casimir force decays algebraicly $\propto k_BT_c\Delta_{(+,+)}/l^d$.
Away from criticality, the critical Casimir force decays exponentially as a function of $l/\xi_\pm$.
The scaling function $k_{\parallel}$ has been calculated exactly in $d=2$ \cite{evans:1994}, for $d \le 4$
using a perturbative field-theoretical method \cite{Krech:1997} or a local-functional method \cite{borjan:2008},  and 
in $d=3$ numerically via Monte Carlo simulations \cite{Vasilyev:2007,Vasilyev:2009,Hasenbusch:2010a,
Hasenbusch:2010,Hasenbusch:2010b}. 
\subsection{Crenellated walls \label{sec:cren}}
%
In the following we consider a crenellated wall located at a minimal surface-to-surface distance $L$ from a
planar wall as sketched in \fref{fig:sketch}. The width and depth of the crenels are given by $W$ and $D$, respectively, 
and the structure is 
periodic along the lateral $x$-direction with period $P$, so that the width of the merlons, i.e., the surface-to-surface 
separation between two neighboring crenels, is given by $P-W$.
Accordingly, the corresponding universal contribution to the critical Casimir force $f$ per area of the planar wall scales as
\begin{equation} 
  \label{eq:cren-force}
  f(L,D,W,P,T)=k_BT \frac{1}{L^d}k(\tau,\delta,\omega,\lambda), 
\end{equation} 
where the geometrical parameters form the following scaling variables:
\begin{equation} 
  \label{eq:def-geo}
  \tau\equiv t\left(\frac{L}{\xi_0^+}\right)^{1/\nu},\quad
  \delta\equiv\frac{D}{L},\quad
  \omega\equiv\frac{W}{L},\quad
  \lambda\equiv\frac{W}{P}\in(0,1).
\end{equation} 
\par
The critical Casimir force between a crenellated and a planar wall attains the value of the corresponding 
force between two planar walls in various limits.
For very shallow crenels, i.e., in the limit $D\ll L$ or $\delta\to0$, the following relation holds:
\begin{equation} 
  \label{eq:derjaguin-limit}
  k(\tau,\delta,\lambda,\omega)\xrightarrow{\delta\to0}k_\parallel(\tau).
\end{equation} 
As will be discussed in more detail below, we find that for narrow crenels with $\omega=W/L\to0$, the order parameter profile 
attains the form of the planar wall geometry, \emph{independent} of the value of $\delta$:
\begin{equation} 
  \label{eq:limit-omega-0}
  k(\tau,\delta,\lambda,\omega)\xrightarrow{\omega\to0}k_\parallel(\tau).
\end{equation} 
Similarly, for $\lambda\to0$, one has
\begin{equation} 
  \label{eq:limit-lambda-0}
  k(\tau,\delta,\lambda,\omega)\xrightarrow{\lambda\to0}k_\parallel(\tau).
\end{equation} 
\par
\begin{figure} 
  \ifTwocolumn
  \includegraphics[width=8.0cm,clip=true]{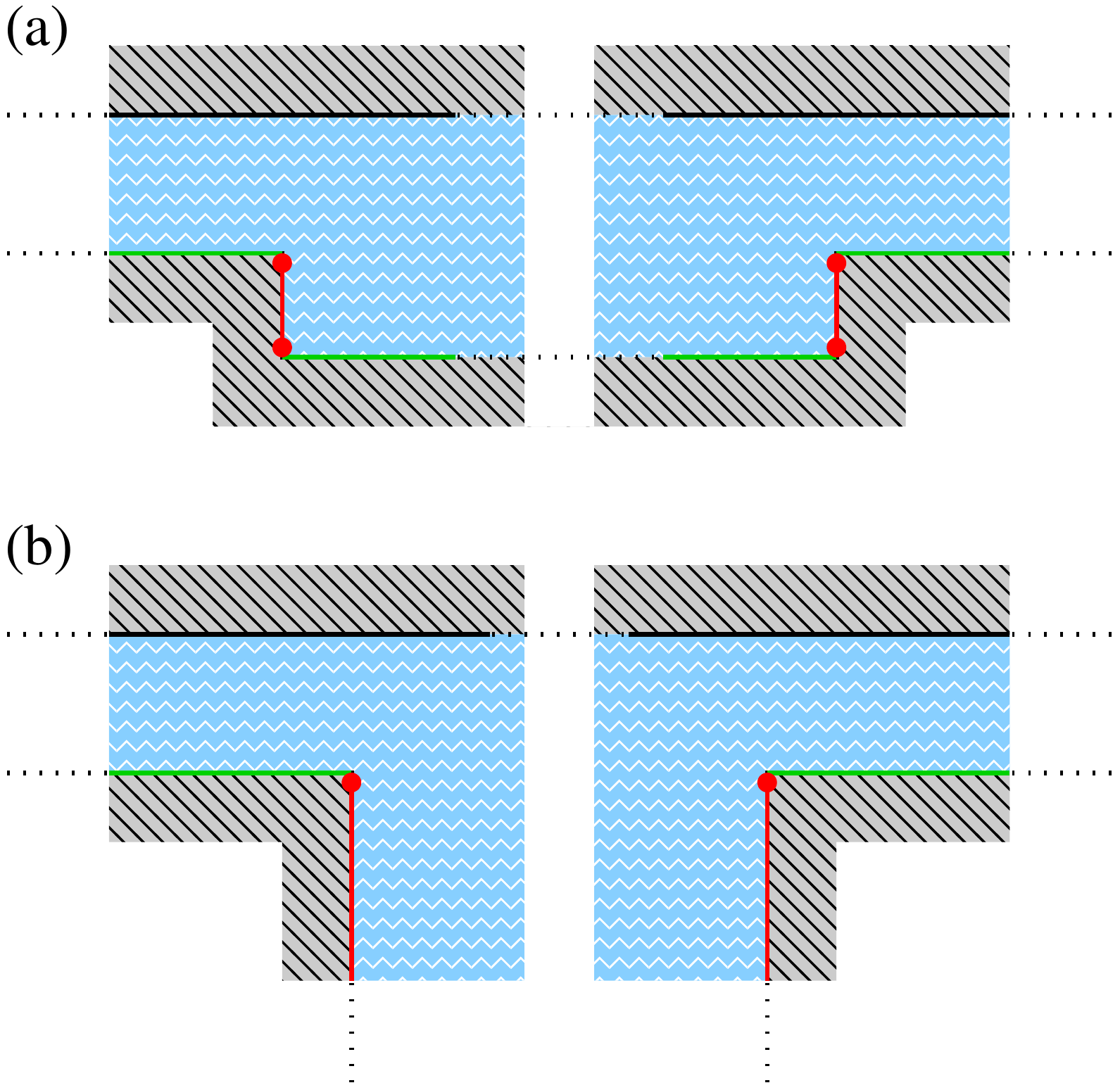} 
  \else
  \includegraphics[width=8.50cm,clip=true]{sketch_step} 
  \fi
  \clevercaption{%
  Sketch concerning the the limiting behavior of the scaling function $k$ (\eref{eq:limit-omega}) 
  for $\omega=W/L\gg1$ at fixed values of $\lambda=W/P$ and $\delta=D/L$. 
  (a) The surfaces of the lower, geometrically structured substrate which are parallel to the film (indicated by a green line) give rise to the 
  PFA contribution $k^\textrm{PFA}$ to the universal scaling function $k$ of the critical Casimir force. In the limit $\omega\to\infty$, 
  the steps consisting of the upper and lower edges as well as the sidewalls (indicated by the red lines and dots) 
  generate the step contribution ${A(\tau,\delta,\lambda)}/{\omega}$ to the critical Casimir force (see \eref{eq:limit-omega}). 
  For fixed $0<\lambda<1$, the step-step interaction vanishes and $A$ becomes independent of $\lambda$. 
  (b) For $\delta\gg1$ the step contribution $A/\omega$ to the critical Casimir force effectively 
  corresponds to the contribution of two right-angled corners opposite to a planar substrate and 
  attains a limiting value independent of $\delta$.
  }   
  \label{fig:sketch_step}
\end{figure}

For very broad crenels compared with the film thickness, i.e., $\omega\to\infty$, the limiting behavior 
of the scaling function of the critical Casimir force is given by the average of the 
scaling functions for two planar walls at distances $L$ and $L+D$, respectively. This corresponds to the 
so-called proximity force approximation (PFA), which we discuss in the following.
Within the PFA the surfaces are subdivided into infinitely small elements parallel to each other, and the resulting
force is obtained by pairwise adding the individual contributions to the force.
The Derjaguin approximation is a special case of the PFA for smoothly
curved surfaces.
Detailed comparisons of the Derjaguin approximation with experimental and theoretical results revealed a good 
agreement for a large range of parameters \cite{Hertlein:2008,Gambassi:2009,Troendle:2011}.
For the present geometry the resulting critical Casimir force $f^{\textrm{PFA}}$ per area [\eref{eq:cren-force}] 
acts along the $z$-direction and is the sum of two contributions: 
(i) the force between the fraction $(1-\lambda)$ of the upper flat wall and the top surfaces of the merlons 
separated by the distance $L$, and (ii) the force between the fraction $\lambda$ of the upper flat wall and 
the bottom surfaces of the crenels separated by the distance $L+D$:
\begin{equation} 
  \label{eq:derjaguin}
  f^\textrm{PFA}(L,D,W,P,T)=(1-\lambda)f_\parallel(L,T)+\lambda f_\parallel(L+D,T).
\end{equation} 
Accordingly, the scaling function of the critical Casimir force within the PFA is given by
\begin{equation} 
  \label{eq:derjaguin-scaling}
  k^\textrm{PFA}(\tau,\delta,\lambda)=(1-\lambda)k_\parallel(\tau)+ \frac{\lambda}{(1+\delta)^d}k_\parallel(\tau(1+\delta)^{1/\nu}).
\end{equation} 
The scaling function $k^\textrm{PFA}$ is independent of $\omega$ because within PFA the effective interactions between the steps and the upper wall are ignored. That is, as long the values of $\lambda$ and $\delta$ are fixed, steps may be arbitrarily added to or removed from the structured substrate (e.g., via a transformation $W\mapsto\alpha W$ and $P\mapsto\alpha P$, where $\alpha>0$) without changing the force within PFA, independent of the value of $\omega$. 
Analogous to the case of chemically striped surfaces, discussed in detail in Refs.~\cite{Parisen:2013,Parisen:2010,Parisen:2014}, every isolated geometrical step  
gives rise to a contribution to the scaling function of the critical Casimir force per area which is proportional to $\omega^{-1}=L/W$. 
The asymptotic behavior for $\omega\to\infty$ of the universal scaling function for the critical Casimir force 
between a planar wall and a crenellated surface is therefore given by
\begin{equation}
  \label{eq:limit-omega}
  k(\tau,\delta,0<\lambda<1,\omega\to\infty)=k^\textrm{PFA}(\tau,\delta,\lambda)+\frac{A(\tau,\delta,\lambda)}{\omega},
\end{equation} 
where we define $A(\tau,\delta,\lambda)$ as the universal contribution of a pair of geometrical steps as sketched in \fref{fig:sketch_step}(a). 
For a fixed value $0<\lambda<1$, in \eref{eq:limit-omega} $A$ actually does not depend on $\lambda$ because in the limit $\omega\to\infty$ the distance $L$ between the steps and the upper wall is much smaller than the step-step distances $W$ and $P-W$, respectively, so that step-step interactions are negligible in this limit. In the following we therefore define $A(\tau,\delta)\equiv A(\tau,\delta,0<\lambda<1)$ and consider this as the generic case. 
For very deep crenels $A(\tau,\delta\to\infty)$ attains a $\tau$-dependent value which corresponds to the contribution to the critical Casimir force between a pair of top 
corners of the right-angled edges of the merlons opposite to a planar wall (see the sketch in \fref{fig:sketch_step}(b)).

\subsection{Mean-field theory \label{sec:mft}}
%
The standard Landau-Ginzburg-Wilson fixed-point effective Hamiltonian describing critical phenomena
of the Ising bulk universality class is given by \cite{Binder:1983,Diehl:1986,Diehl:1997}
\begin{equation} 
  \label{eq:hamiltonian}
  \mathcal{H}[\phi]=\int_V\,\upd^d{r}\,\left\{
        \frac{1}{2}(\nabla\phi)^2
        +\frac{{\hat{\tau}}}{2}\phi^2
       +\frac{u}{4!}\phi^4
       -h\phi
			 \right\},
\end{equation} 
where $\phi(\vec{r})$ is proportional to the order parameter describing the fluid, which completely fills the
accessible volume $V$ in $d$-dimensional space.
The statistical weight of a configuration  $\phi(\vec{r})$ is proportional to $\exp(-\mathcal{H})$.
The parameter ${\hat{\tau}}$ in \eref{eq:hamiltonian} is proportional to $t$, and $u>0$ is a coupling constant.
The last term in \eref{eq:hamiltonian} vanishes for the case considered here (i.e., $h=0$), which 
corresponds to the situation that the concentrations of the species forming the binary liquid mixture are fixed to their
critical values.
In a finite-size system the bulk Hamiltonian $\mathcal{H}[\phi]$ is supplemented by appropriate 
surface and curvature (edge) contributions \cite{Binder:1972,Binder:1983,Diehl:1986,Diehl:1997}.
This surface contribution, which adds to \eref{eq:hamiltonian},
is given by \cite{Binder:1972,Binder:1983,Diehl:1986,Diehl:1997}
\begin{equation} 
  \label{eq:hamil-surf}
  \mathcal{H}_{\rm s}[\phi]=\int_{\partial V}\,\upd^{(d-1)}{r}\,\left\{
            \frac{c}{2}\phi^2
            -h_1\phi
             \right\},
\end{equation}
where $c$ is the so-called surface enhancement and $h_1$ is a surface field; $\partial V$ is the surface of the volume $V$.
In the strong adsorption limit \cite{burkhardt:1994,Diehl:1993}, as discussed in the present study, these contributions 
generate boundary conditions for the order parameter such that $\phi\big|_{\text{surface}}=\infty$ corresponding
to $(+)$ BCs.
Thus, the use of \eref{eq:hamil-surf} together with additional surface contributions can be replaced by applying
the appropriate BC to $\phi$ and by using \eref{eq:hamiltonian} throughout the bulk.
The mean-field order parameter profile minimizes
the Hamiltonian, i.e., $\updelta \mathcal{H}[\phi]/\updelta\phi|_{\phi=\langle\phi\rangle}=0$.
In the \emph{bulk}, the mean-field order parameter is spatially constant and attains the values
$\langle\phi\rangle_b=\pm B|\temp|^\beta$ for $\temp<0$ and $\langle\phi\rangle=0$ for $\temp>0$, where, besides $\xi_0^+$,
$B$ is the only additional independent nonuniversal amplitude appearing 
in the description of bulk critical phenomena; $\beta(d=4)=1/2$ is a standard critical exponent.
Within mean-field theory (MFT) the following relations hold: ${\hat{\tau}}=t (\xi_0^+)^{-2}$ and $u=6B^{-2}(\xi_0^+)^{-2}$.
\par
For the film geometry, the MFT scaling function for the critical Casimir force can be determined analytically 
\cite{Krech:1997}. One finds [see \eref{eq:delta-ab}] for the case of the same strong adsorption at both surfaces
the critical Casimir amplitude $\Dpp=24[K(1/\sqrt{2})]^4/u\simeq-283.61\times u^{-1}$, 
where $K$ is the complete elliptic integral of the first kind \cite{Krech:1997}.
Renormalization group arguments tell that MFT provides the correct universal properties of critical phenomena 
for spatial dimensions above the upper critical dimension, i.e., $d \ge d_{uc} =4$, up to logarithmic corrections 
in $d=d_{uc}$ \cite{zinn-justin:2002}. Moreover, MFT provides the lowest-order contribution to universal quantities 
within an expansion in terms of $\epsilon = 4 - d$.

\section{Order parameter profiles \label{sec:profiles}}
%
\begin{figure} 
  \ifTwocolumn
  \includegraphics[width=8.5cm,clip=true,trim=0 0 12 34]{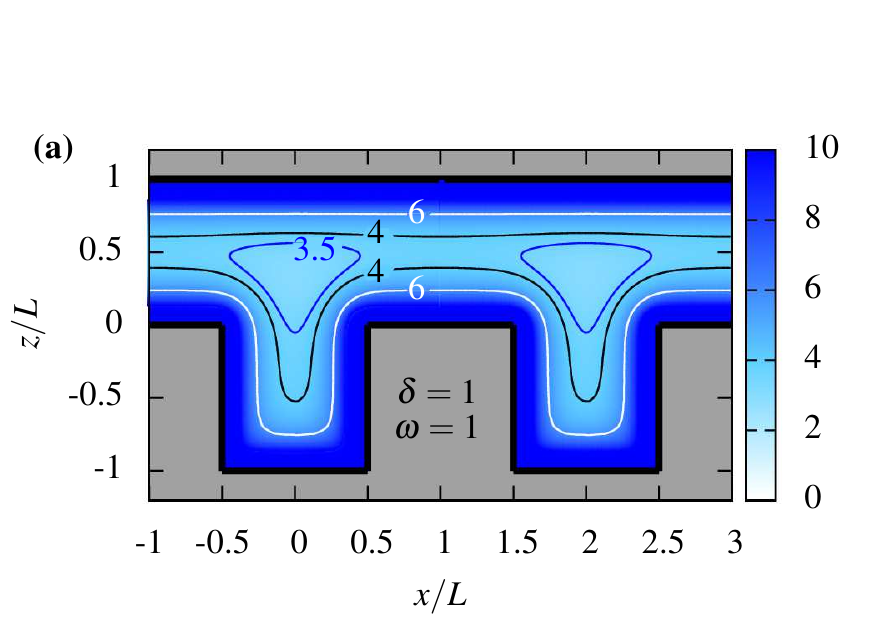} 
  \includegraphics[width=8.5cm,clip=true,trim=0 0 12 24]{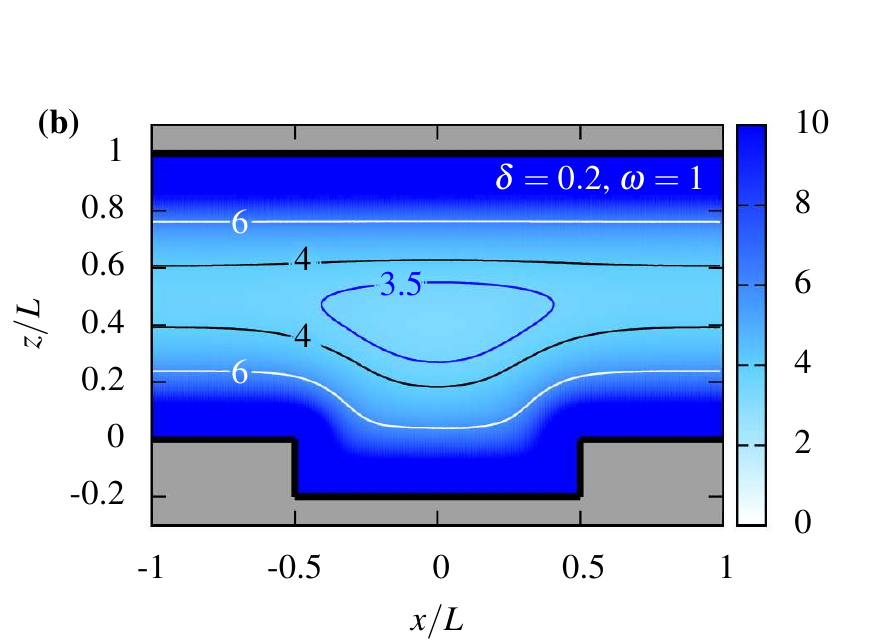} 
  \includegraphics[width=8.5cm,clip=true,trim=0 0 12 34]{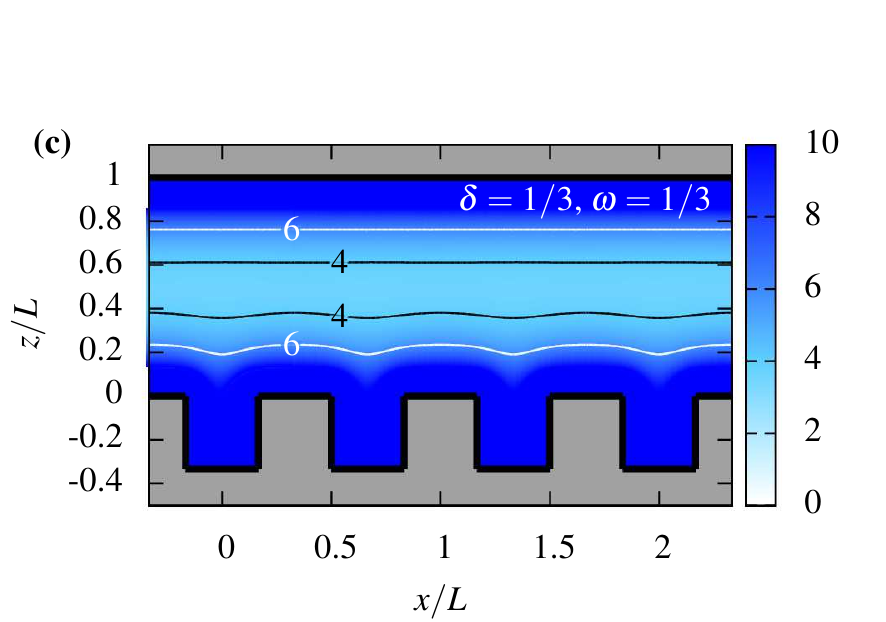} 
  \else
  \includegraphics[width=8.50cm,clip=true,trim=0 0 12 34]{plot1a} 
  \includegraphics[width=8.50cm,clip=true,trim=0 0 12 24]{plot2a} 
  \includegraphics[width=8.50cm,clip=true,trim=0 0 12 34]{plot3a} 
  \fi
  \clevercaption{%
  Rescaled MFT order parameter profile $m(\vec{r})\times L$ (see Eqs. \eqref{eq:mftop} and \eqref{eq:mftop-2}) at $T=T_c$ (i.e., $\tau=0$) for
  various values of $\delta=D/L$ and $\omega=W/L$ with $\lambda=1/2$ fixed (see \fref{fig:sketch}):
  (a) $\delta=\omega=1$ (deep crenels and widely spaced merlons), (b) 
  $\delta=0.2$, $\omega=1$ (shallow crenels and widely spaced merlons), and (c) 
  and $\delta=\omega=1/3$ (shallow crenels and closely spaced merlons).
  The local values of $m(\vec{r})\times L$ are indicated by the color code given by the side bars.
  Large values of $m\times L\gg1$ indicate strong adsorption of the fluid close 
  to the surfaces.
  From the shape of the contour lines for $m\times L=6$ near the structured wall (lower white lines) the dependence
  of the critical adsorption profile on the scaling variables $\delta$ and $\omega$ is clearly visible.
  Whereas for widely spaced merlons with $\omega=1$ the adsorption profile follows the shape of the crenelated 
  surface, for closely spaced merlons with $\omega=1/3$
  the crenels are filled with the adsorbed fluid and for increasing normal distances $z>0$ the order parameter profile rapidly adopts 
  an effective planar wall behavior corresponding to rather straight contour lines.
  }   
  \label{fig:op}
\end{figure}
The order parameter $\phi$ exhibits the following scaling properties (see Subsec. 2.5 in Ref.~\cite{Parisen:2010}):
\begin{equation}
  \phi(t,x,z,L;D,W,P)=B|t|^\beta Q_\pm\left(\tfrac{x}{\xi_\pm},\tfrac{z}{\xi_\pm},\tfrac{L}{\xi_\pm};\delta,\omega,\lambda\right)
\end{equation}
or equivalently
\begin{equation} 
  \phi(t,x,z,L;D,W,P)=B\left(\frac{L}{\xi_0^+}\right)^{-\beta/\nu} R_\pm\left(\tfrac{x}{L},\tfrac{z}{L},\tau;\delta,\omega,\lambda\right)
\end{equation} 
with universal scaling functions $Q_\pm$ and $R_\pm$. The bulk order parameter varies as $\phi_b=B|t|^\beta$. This implies
\begin{equation} 
  \label{eq:relation}
  \left(\frac{L}{\xi_0^+}\right)^{\beta/\nu}\frac{1}{B}\phi(\vec{r})=R_\pm\left(\tfrac{x}{L},\tfrac{z}{L},\tau;\delta,\omega,\lambda\right).
\end{equation} 
Within MFT one has $\beta=\nu=1/2$ so that with the definition
\begin{equation} 
  \label{eq:mftop}
  m(\vec{r})\equiv {(B\xi_0^+)^{-1}}{\langle\phi(\vec{r})\rangle}
\end{equation} 
\eref{eq:relation} renders the following MFT approximation for the scaling function $R$:
\begin{equation} 
  \label{eq:mftop-2}
  m(\vec{r})\times L=R^{\textrm{MFT}}\left(\tfrac{x}{L},\tfrac{z}{L},\tau;\delta,\omega,\lambda\right).
\end{equation}

\par
In the following we present these MFT results which we have obtained by minimizing numerically $\mathcal{H}[\phi]$ 
using a finite element method in order to obtain the (spatially inhomogeneous) profile $m(\vec{r})$ for the geometries under consideration.
Here, we focus on the case of strong adsorption and the same chemical BCs at the two surfaces.
For distances from the surface of a substrate which are small compared to $\xi_\pm$, or for 
$T \to T_c$, the order parameter varies algebraicly. In order to obtain a BC 
for the numerical calculations we use a short distance expansion (see, e.g., Ref.~\cite{Troendle:2008} 
and reference therein) with $m=\infty$ at the surfaces of the two walls shown in 
Fig.~\ref{fig:sketch}.
\par
In \fref{fig:op} the order parameter profile of a fluid confined between a planar and a crenellated wall
at the bulk critical point $T=T_c$ is shown for $\lambda= W/P =0.5$; $m(\vec{r})$ depends on $x$ 
and $z$ and is invariant along the $y$-direction.
From \fref{fig:op} we can infer that for $\omega=W/L=1$ the order parameter profile follows the 
shape of the crenellated surface [Figs. \ref{fig:op} (a) and (b)], whereas for $\omega=1/3$ the space 
between the merlons corresponds to high values of the order parameter, i.e., the crenels are ``filled'' with
the adsorbed fluid [\fref{fig:op} (c)] until an almost straight contour line has formed 
separating the fluid with high order parameter, which fills the crenels, from the fluid with lower 
order parameter in the middle of the slit. Hence for narrow crenels, for 
increasing $z>0$ the order parameter profiles rapidly approach the ones of a corresponding 
film of thickness $L$.

\section{Scaling function of the critical Casimir force \label{sec:force}}
%
The critical Casimir forces are calculated directly from the numerically obtained mean-field order parameter
profiles using the stress tensor \cite{Krech:1997,Kondrat:2009}. 
As in Sec.~\ref{sec:profiles}, here throughout we focus on the case $\lambda=W/P=0.5$.
We estimate the numerical error of the present method to be less than 1\%.
\par
\begin{figure} 
  \ifTwocolumn
  \includegraphics[width=8.0cm,clip=true]{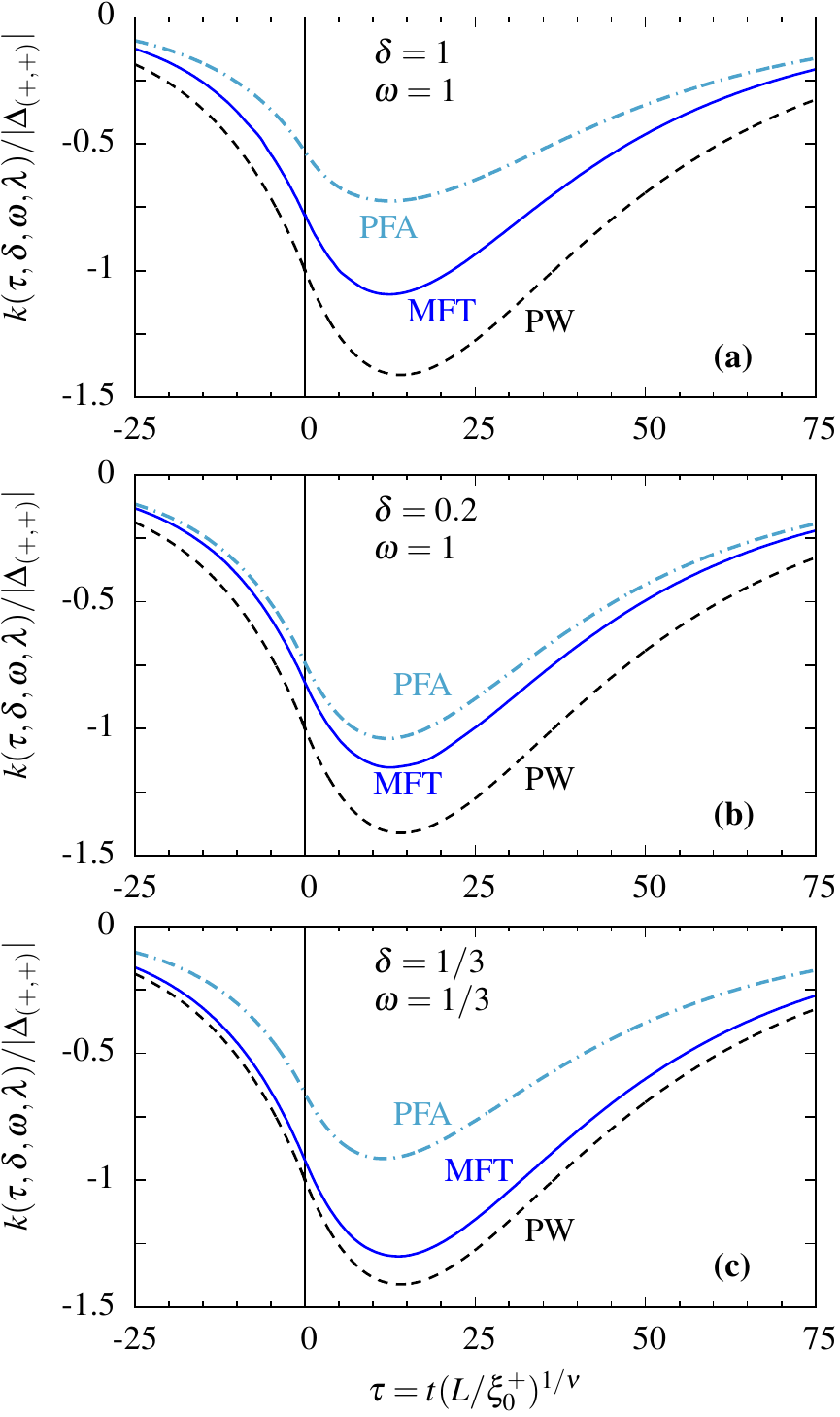} 
  \else
  \includegraphics[width=8.50cm,clip=true]{force1} 
  \fi
  \clevercaption{%
  Normalized scaling function $k(\tau,\delta,\omega,\lambda)/|\Dpp|$ of the critical Casimir force as a function of $\tau=t(L/\xi_0^+)^{1/\nu}$ for a
  fixed ratio $\lambda=W/P=0.5$ and various values of $\delta=D/L$ and $\omega=W/L$.
  In addition to the numerically obtained full MFT scaling functions (full curves), the corresponding scaling functions
  $k^\textrm{PFA}$ obtained within the PFA [\eref{eq:derjaguin-scaling}] (dashed curves) and the
  scaling function for two planar walls (PW) at distance $L$ (dashed-dotted curves) are shown.
  For $\delta=\omega=1$, in (a) the MFT scaling function does neither agree with the one obtained within PFA nor with the one for the PW case, but 
  lies roughly in between these two curves.
  However, for shallow crenels and widely spaced merlons with $\delta=0.2$ and $\omega=1$, in (b) the PFA
  is closer to the full MFT scaling function.
  For shallow crenels and closely spaced merlons with $\delta=\omega=1/3$, in (c) the full MFT scaling function is 
  similar to the one for two planar walls (PW).
  Figures \ref{fig:op} and \ref{fig:force} allow a direct comparison between the order parameter distribution and the 
  critical Casimir force at $T_c$, i.e., for $\tau = 0$.
  Here the scaling function $k$ is normalized with $|\Dpp|$ so that the ratio does not depend on the coupling constant $u$, which within MFT is undetermined.
  }   
  \label{fig:force}
\end{figure}
In \fref{fig:force} the scaling function $k(\tau,\delta,\omega,\lambda)$ of the critical Casimir force 
is shown as a function of $\tau=t(L/\xi_0^+)^{1/\nu}$ for various values of $\delta=D/L$ and $\omega=W/L$.
In addition to the numerically obtained full MFT scaling functions (solid curves), the corresponding scaling functions
$k^\textrm{PFA}(\tau,\delta,\omega,\lambda)$ obtained within the PFA [\eref{eq:derjaguin-scaling}] (dashed curves), 
and the scaling function $k_\parallel(\tau)$ for two planar walls (PW) at distance $L$ (dashed-dotted curves) are shown.
For $\delta=\omega=1$, in \fref{fig:force}(a) the MFT scaling function does neither agree with the one obtained within the 
PFA nor with the one for the PW case, but lies roughly in between these two curves. The corresponding MFT order parameter profile at $T_c$ is shown in 
\fref{fig:op}(a).
For $\delta=0.2$ and $\omega=1$, in \fref{fig:force}(b) the scaling function obtained within the PFA is closer to the 
full MFT scaling function which corresponds to the critical MFT order parameter profile shown in \fref{fig:op}(b).
For these shallow crenels and widely spaced merlons the relative contribution from the the right-angled steps of the merlons to the total critical Casimir force 
is smaller than for the case shown in \fref{fig:force}(a).
For shallow crenels and closely spaced merlons with $\delta=\omega=1/3$, in \fref{fig:force}(c) the MFT scaling function is 
similar to the one for two planar walls (PW). This corresponds to the case in which the crenels are filled by a fluid with a high value of the order
parameter (see \fref{fig:op}(a) for $T=T_c$).
\par
As can be inferred from \fref{fig:force}, in general the PFA deviates from the full MFT results because critical phenomena do not allow for linear superposition.
In order to study the deviations of the results for the force from the corresponding ones following from the assumption of pairwise additivity, we study the ratio
$k/k^{\textrm{PFA}}$ of the scaling function $k$ obtained within full MFT and the one ($k^{\textrm{PFA}}$) obtained within PFA. 
From Eqs.~(\ref{eq:limit-omega-0}) and (\ref{eq:derjaguin-scaling}) we find for $\tau=0$, i.e., $T=T_c$:
\begin{equation} 
  \label{eq:limit-omega-paa}
  \frac{k(\tau=0,\delta,\lambda,\omega)}{k^{\textrm{PFA}}(\tau=0,\delta,\lambda)}\xrightarrow{\omega\to0}
  \frac{1}{1-\lambda+ {\lambda}{(1+\delta)^{-d}}}.
\end{equation} 

\par

\begin{figure} 
  \ifTwocolumn
  \includegraphics[width=8.0cm,clip=true]{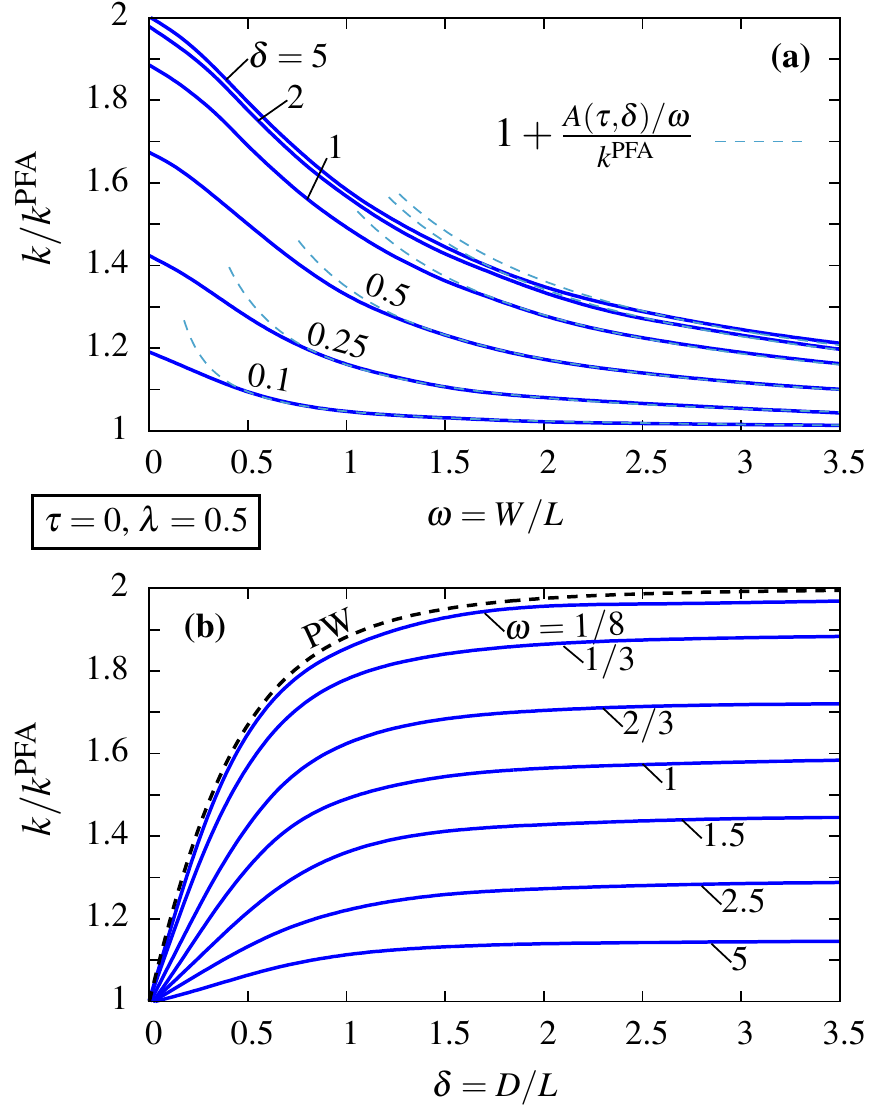} 
  \else
  \includegraphics[width=8.50cm,clip=true]{plot_o_d_final} 
  \fi
  \clevercaption{%
  Ratio $k/k^{\textrm{PFA}}$ of the scaling function $k$ of the critical Casimir force obtained within full MFT and the 
  one ($k^{\textrm{PFA}}$) obtained within PFA, for $\tau=0$ (i.e., $T=T_c$) and $\lambda=0.5$ (\eref{eq:limit-omega-paa}). 
  The ratio varies between $1$ and $(1-\lambda)^{-1}=2$ corresponding to deviations of the actual
  force from the force calculated within PFA between 0\% and 100\%, respectively.
  In (a) $k/k^{\textrm{PFA}}$ is shown as a function of $\omega=W/L$ for several values of $\delta=D/L$.
  For widely spaced merlons ($\omega\gg1$),  according to \eref{eq:limit-omega} the ratio $k/k^{\textrm{PFA}}$ approaches $1$. 
  The corresponding decay
  $\propto A(\tau=0,\delta)/\omega$ is shown as dashed lines.
  For very shallow crenels ($\delta\to0$), the ratio $k/k^{\textrm{PFA}}$ approaches $1$, so that PFA is valid.
  In (b) $k/k^{\textrm{PFA}}$ is shown as a function of $\delta=D/L$ for various values of $\omega$. 
  We find that, for deep crenels ($\delta\gg1$), $k/k^{\textrm{PFA}}$ attains a plateau.
  For large values of $\omega$ this corresponds to the limiting behavior given in \eref{eq:limit-omega};
  ultimately, for $\omega\to\infty$ PFA is valid.
  On the other hand for closely spaced merlons ($\omega\to0$) the critical Casimir force approaches its value for an effective film 
  geometry with the two parallel walls at distance $L$. 
  In this limit the ratio $k/k^{\textrm{PFA}}$ is given by \eref{eq:limit-omega-paa} with $\lambda=0.5$ and $d=4$ (dashed line, ``PW'').
  }   
  \label{fig:compare}
\end{figure}
Figure~\ref{fig:compare} shows the ratio $k/k^{\textrm{PFA}}$ as function of $\delta$ and $\omega$ for fixed values 
$\tau=0$ and $\lambda=0.5$ (\eref{eq:limit-omega-paa}).
The ratio $k/k^{\textrm{PFA}}$ varies between $1$ and $(1-\lambda)^{-1}=2$ which corresponds to deviations of the 
actual force from the one calculated within PFA between 0\% and 100\%, respectively.
In \fref{fig:compare}(a) $k/k^{\textrm{PFA}}$ is shown as a function of $\omega$ for various values of $\delta$. This graph tells that the limiting behavior for $\omega\to\infty$
given in \eref{eq:limit-omega} already holds for $\omega\gtrsim2$.
The amplitude function $A(\tau=0,\delta)$ [\eref{eq:limit-omega}] has been determined via a 
least square fit to the numerical data.
As expected on physical grounds, for very shallow crenels the critical Casimir force 
can be approximated by the corresponding PFA expression and hence $k/k^{\textrm{PFA}}\to1$ for $\delta\to0$.
In \fref{fig:compare}(b) we show $k/k^{\textrm{PFA}}$ as a function of $\delta$ for various values of $\omega$. 
For deep crenels with $\delta\gtrsim2$ this ratio reaches a plateau, i.e., for $\delta\gg1$ the strength of the critical
Casimir force is not affected by the depth of the crenels so that for $\omega\to\infty$ PFA becomes valid.
For $\omega\to0$ the critical Casimir force approaches its value for an effective film geometry with the two parallel
walls (dashed line, ``PW'') at distance $L$ (see \eref{eq:limit-omega-paa} for $\lambda=0.5$ and $d=4$).
This means that for closely spaced merlons the critical Casimir force reduces to that between two parallel flat surfaces at separation $L$.
\par
\begin{figure} 
  \ifTwocolumn
  \includegraphics[width=8.0cm]{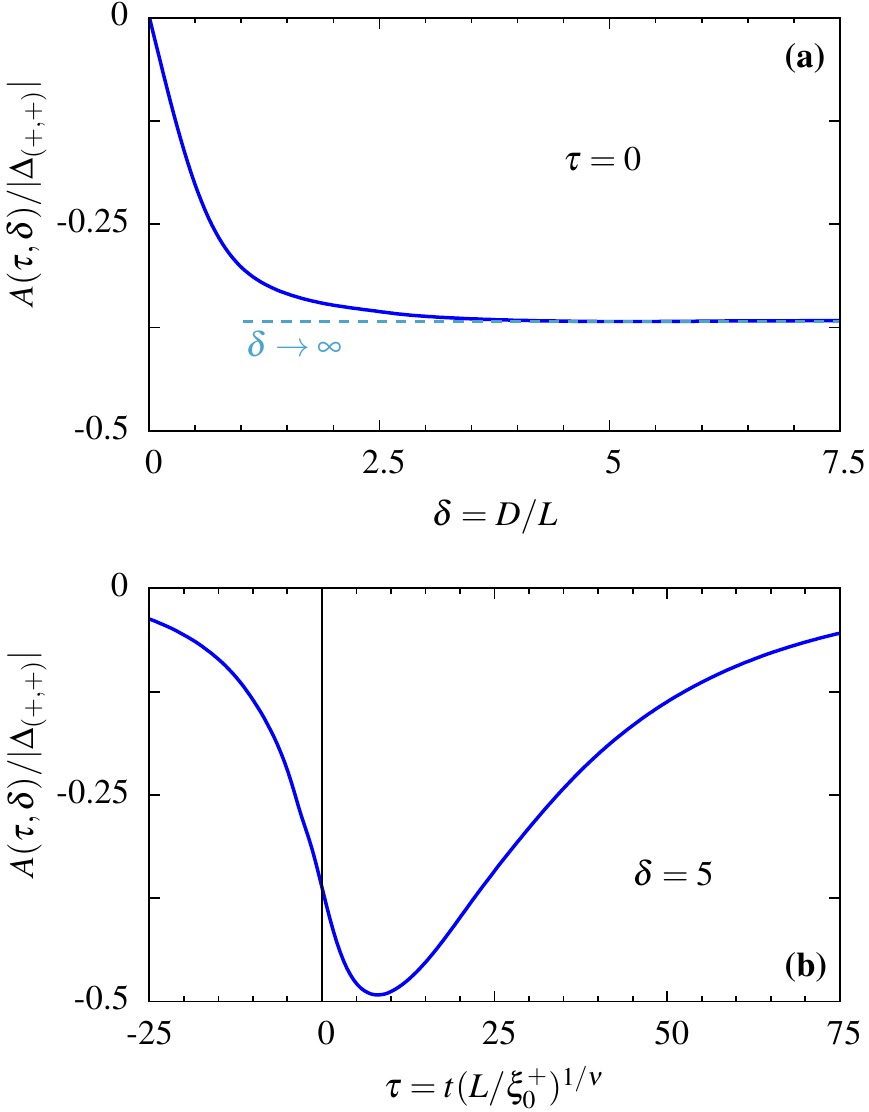} 
  \else
  \includegraphics[width=8.50cm]{a_delta} 
  \fi
  \clevercaption{%
  (a) Normalized amplitude $A(\tau,\delta)/|\Dpp|$ [see \eref{eq:limit-omega}] of the step contribution to the critical Casimir force for
  $\tau=0$ as a function of the reduced depth $\delta=D/L$  of the crenels. For $\delta\gtrsim5$ a plateau 
  value $A(\tau=0,\delta\to\infty)/|\Dpp|\simeq-0.367$ is reached and the step contribution to the critical Casimir force becomes independent 
  of the depth of the crenels.
  In (b) $A(\tau,\delta)/|\Dpp|$ is shown as a function of $\tau$ as obtained for $\delta=5$. 
  According to our analysis, for $\delta=5$ the
  limiting behavior for $\delta\to\infty$ is almost reached (see (a)). (According to \eref{eq:limit-omega} $A$ does not depend on $\lambda$ as long as the value of the latter is not $0$ or $1$.)
  }   
  \label{fig:A}
\end{figure}
As can be inferred from \eref{eq:limit-omega} and \fref{fig:compare}(a) for large values of $\omega$
the critical Casimir force acting between a flat and a crenellated wall can be described reliably by the sum of 
the PFA-contribution and a contribution $A(\tau,\delta)/\omega$.
The latter contribution stems from effects due to the presence of geometrical steps (see \fref{fig:sketch_step}).
In \fref{fig:A}(a) we show $A(\tau,\delta)$ as a function of $\delta$ 
as obtained numerically within full MFT via least square fits to the data
for $\tau=0$.
As expected, for $\delta\to0$, i.e., $D/L\to0$, $A(\tau,\delta)$ vanishes for the case
of chemically homogeneous boundary conditions as considered here. 
(For chemically inhomogeneous boundary conditions in lateral direction a line contribution arises due to the presence
of a \emph{chemical} step even for a planar substrate \cite{Parisen:2013,Parisen:2010,Parisen:2014}.)
For $\delta\gg1$, i.e., deep crenels, $A(\tau,\delta)$ attains negative values and depends on $\tau$ only.
$A/\omega$ corresponds to the contribution to the critical Casimir force of two infinitely extended right-angled corners 
opposite to a planar substrate (see \fref{fig:sketch_step}(b)).
We find that this plateau value is reached for $\delta\gtrsim5$. 
In this limit the lower parts of the crenels including the right-angled wedges at the bottom do not contribute to the critical Casimir force.

Figure~\ref{fig:A}(b) shows the amplitude $A(\tau,\delta)$ of the step contribution as a function of $\tau$ for 
deep crenels with $\delta=5$, normalized by the absolute value $|\Dpp|$ of the critical Casimir amplitude.
For the chosen value $\delta=5$ we find that $k^{\textrm{PFA}}(\tau,\delta\gg1,\lambda)\simeq (1-\lambda) k_\parallel(\tau)$
holds [see \eref{eq:derjaguin-scaling}].
We find that the functional shapes of this amplitude of the step contribution  and of $ k_\parallel(\tau)$ are similar but
the minima of the two scaling functions are displaced such that the minimum of $A$ as a function of
$\tau$ is located closer to the critical point at $\tau=0$. For comparison we note that the functional 
shape of the amplitude of the step contribution is very close to the numerically obtained MFT scaling function of the attractive critical Casimir force 
acting between a thin cylinder and a planar substrate, where the symmetry axis of the cylinder is 
parallel to the substrate \cite{Labbe:2014}. Indeed, the lower white contour line in Fig.~\ref{fig:op}(a) 
indicates that one may approximate each cross-section of the two right-angled edges of the merlons 
by a small inscribed quarter circle, keeping in mind that in the limit $\xi_\pm/R\to\infty$ the critical adsorption profile near 
a thin cylinder is independent of the radius $R$ of the cylinder \cite{Hanke:1999a}.

\section{Conclusions \label{sec:summary}}

Within mean field theory (MFT) we have calculated the critical Casimir force between a planar and a crenellated surface,
i.e., a periodic structure of geometric steps forming crenels and merlons (\fref{fig:sketch}).
To this end we have numerically calculated the critical order parameter profiles of a fluid confined by such a structure.
We have focused on the experimentally relevant case of binary liquid mixtures or simple liquids, which belong
to the Ising bulk universality class, and on $(+)$ boundary conditions (BCs) which correspond to the generic
case of strong critical adsorption.
Using the stress tensor we have calculated the universal scaling function of the singular contribution to
the critical Casimir force, acting on the confining walls along the normal direction upon approaching the critical point.
We have obtained the following main results:
\begin{enumerate}
  \item 
    According to finite size scaling, the universal scaling function of the critical Casimir force
    depends on only a few scaling variables which describe the geometry under consideration 
    [Eqs.~\eqref{eq:cren-force}, \eqref{eq:def-geo}, and \fref{fig:sketch}].
    In the limits, that the width $W$ or the depth $D$ of the crenels are small compared
    to the film width or the period of the geometric structure, the force reduces to the one acting between two 
    parallel flat substrates at a distance $L$
    [Eqs.~\eqref{eq:derjaguin-limit} -- \eqref{eq:limit-lambda-0}].
    On the other hand, for fixed reduced crenel depth $\delta$ and for reduced crenel width $\omega=W/l\to\infty$, 
    the expression for the critical Casimir force approaches the one obtained
    via the proximity force approximation (PFA) [\eref{eq:derjaguin}] 
    as $A(\tau,\delta,\lambda)/\omega$, which can be interpreted as a step contribution [see \eref{eq:limit-omega} and \fref{fig:sketch_step}].

  \item
    In Sec.~\ref{sec:profiles} we have shown generic examples of order parameter profiles (i.e.,
    the universal contribution to the profiles of the deviations of the local concentration or density from its critical value in the bulk) as obtained within MFT 
    at the critical point.
    Whereas for crenel sizes comparable with the film thickness the order parameter profile follows
    the geometrical structure [Figs.~\ref{fig:op}(a) and (b)], for shallow crenels and closely spaced merlons
    the order parameter profile resembles the one in between two planar walls [\fref{fig:op}(c)].
    In this latter case strong critical adsorption
    effectively suppresses the influence of the merlons.

  \item
    These properties of the order parameter profiles agree with our results for the universal scaling function $k$ [\eref{eq:cren-force}] of the critical
    Casimir force presented in Section~\ref{sec:force}:
    For deep crenels and widely spaced merlons the full MFT scaling function of the critical
    Casimir force as a function of the scaling variable $\tau$ does neither agree with the planar wall limit 
    (PW) nor with the PFA [\fref{fig:force}(a)]. 
    On the other hand, for shallow crenels and widely spaced merlons the PFA describes the actual full MFT behavior better
    [\fref{fig:force}(b)], whereas for shallow crenels and closely spaced merlons the PW limit is approached [\fref{fig:force}(c)].
    Thus, depending on the values of the scaling variables $\delta=D/L$ and $\omega=W/L$, the critical Casimir force interpolates
    between its limiting behaviors described by the PFA, which always underestimates the strength of the force, and
    the PW limit, which overestimates the critical Casimir force.
    This can also be seen in \fref{fig:compare}(b) which shows the ratio of the universal scaling function of the critical Casimir force 
    and its corresponding PFA value as a function of $\delta$.

  \item
    Upon increasing $\omega$ the scaling function $k^\textrm{PFA}$ as obtained within PFA is approached as $A(\tau,\delta,\lambda)/\omega$ [\fref{fig:compare}a)], 
    which can be interpreted as a step contribution with $A(\tau,\delta)\equiv A(\tau,\delta,0<\lambda<1)$ de facto 
    independent of $\lambda$ [\eref{eq:limit-omega}].
    Thus, for widely spaced merlons the critical Casimir force can be described as
    the linear superposition of the planar wall contributions at distances $L$ and $L+D$, respectively, plus additional step contributions
    stemming from the geometrical steps.
    For large values of $\delta$ this step contribution approaches a specific negative value, independent of the
    crenel depth $D$ [\fref{fig:A}(a)].

\end{enumerate}

To conclude, our numerical calculations within MFT extend previous investigations of the critical Casimir force
caused by the confinement due to structured substrates.
In particular, we have analyzed in detail the crossover of the universal scaling function of the critical Casimir force from an effective
planar wall limit for small roughness to the PFA limit.
This may not only be useful for the analysis of critical Casimir forces in such experimentally relevant, designed geometries, 
but could also help in understanding unavoidable roughness corrections in the case of planar geometries.
Moreover, our results show that the first-order correction to the PFA can be expressed in terms of a
contribution stemming from the individual geometrical steps.
This is analogous to similar situations involving chemically structured surfaces \cite{Parisen:2013}, 
which are discussed in Ref. \cite{Parisen:2014}.
For comparison we note that such step contributions arising from crenellated surfaces are also of importance 
for non-critical fluids consisting of rod-like particles close to the isotropic to nematic phase transition 
of the bulk fluid \cite{harnau:2004}.

\section*{Acknowledgments}
The authors thank Nikolas Brunner for very helpful contributions.


\begin{thebibliography}{77}%

\makeatletter
\providecommand \@ifxundefined [1]{%
 \@ifx{#1\undefined}
}%
\providecommand \@ifnum [1]{%
 \ifnum #1\expandafter \@firstoftwo
 \else \expandafter \@secondoftwo
 \fi
}%
\providecommand \@ifx [1]{%
 \ifx #1\expandafter \@firstoftwo
 \else \expandafter \@secondoftwo
 \fi
}%
\providecommand \natexlab [1]{#1}%
\providecommand \enquote  [1]{``#1''}%
\providecommand \bibnamefont  [1]{#1}%
\providecommand \bibfnamefont [1]{#1}%
\providecommand \citenamefont [1]{#1}%
\providecommand \href@noop [0]{\@secondoftwo}%
\providecommand \href [0]{\begingroup \@sanitize@url \@href}%
\providecommand \@href[1]{\@@startlink{#1}\@@href}%
\providecommand \@@href[1]{\endgroup#1\@@endlink}%
\providecommand \@sanitize@url [0]{\catcode `\\12\catcode `\$12\catcode
  `\&12\catcode `\#12\catcode `\^12\catcode `\_12\catcode `\%12\relax}%
\providecommand \@@startlink[1]{}%
\providecommand \@@endlink[0]{}%
\providecommand \url  [0]{\begingroup\@sanitize@url \@url }%
\providecommand \@url [1]{\endgroup\@href {#1}{\urlprefix }}%
\providecommand \urlprefix  [0]{URL }%
\providecommand \Eprint [0]{\href }%
\providecommand \doibase [0]{http://dx.doi.org/}%
\providecommand \selectlanguage [0]{\@gobble}%
\providecommand \bibinfo  [0]{\@secondoftwo}%
\providecommand \bibfield  [0]{\@secondoftwo}%
\providecommand \translation [1]{[#1]}%
\providecommand \BibitemOpen [0]{}%
\providecommand \bibitemStop [0]{}%
\providecommand \bibitemNoStop [0]{.\EOS\space}%
\providecommand \EOS [0]{\spacefactor3000\relax}%
\providecommand \BibitemShut  [1]{\csname bibitem#1\endcsname}%
\let\auto@bib@innerbib\@empty
\bibitem [{\citenamefont {Casimir}(1948)}]{Casimir:1948}%
  \BibitemOpen
  \bibfield  {author} {\bibinfo {author} {\bibfnamefont {H.~G.~B.}\
  \bibnamefont {Casimir}},\ }\href@noop {} {\bibfield  {journal} {\bibinfo
  {journal} {Proc. K. Ned. Akad. Wet.}\ }\textbf {\bibinfo {Volume} {51}},\
  \bibinfo {pages} {793} (\bibinfo {year} {1948})}\BibitemShut {NoStop}%
\bibitem [{\citenamefont {Capasso}\ \emph {et~al.}(2007)\citenamefont
  {Capasso}, \citenamefont {Munday}, \citenamefont {Iannuzzi},\ and\
  \citenamefont {Chan}}]{Capasso:2007}%
  \BibitemOpen
  \bibfield  {author} {\bibinfo {author} {\bibfnamefont {F.}~\bibnamefont
  {Capasso}}, \bibinfo {author} {\bibfnamefont {J.~N.}\ \bibnamefont {Munday}},
  \bibinfo {author} {\bibfnamefont {D.}~\bibnamefont {Iannuzzi}}, \ and\
  \bibinfo {author} {\bibfnamefont {H.~B.}\ \bibnamefont {Chan}},\ }\href@noop
  {} {\bibfield  {journal} {\bibinfo  {journal} {IEEE J. Quantum Electron.}\
  }\textbf {\bibinfo {Volume} {13}},\ \bibinfo {pages} {400} (\bibinfo {year}
  {2007})}\BibitemShut {NoStop}%
\bibitem [{\citenamefont {Fisher}\ and\ \citenamefont
  {de~Gennes}(1978)}]{Fisher:1978}%
  \BibitemOpen
  \bibfield  {author} {\bibinfo {author} {\bibfnamefont {M.~E.}\ \bibnamefont
  {Fisher}}\ and\ \bibinfo {author} {\bibfnamefont {P.~G.}\ \bibnamefont
  {de~Gennes}},\ }\href@noop {} {\bibfield  {journal} {\bibinfo  {journal} {C.
  R. Acad. Sci., Paris, Ser. B}\ }\textbf {\bibinfo {Volume} {287}},\ \bibinfo
  {pages} {207} (\bibinfo {year} {1978})}\BibitemShut {NoStop}%
\bibitem [{\citenamefont {Krech}(1994)}]{Krech:book}%
  \BibitemOpen
  \bibfield  {author} {\bibinfo {author} {\bibfnamefont {M.}~\bibnamefont
  {Krech}},\ }\href@noop {} {\emph {\bibinfo {title} {The Casimir Effect in
  Critical Systems}}}\ (\bibinfo  {publisher} {World Scientific, Singapore},\
  \bibinfo {year} {1994})\BibitemShut {NoStop}%
\bibitem [{\citenamefont {Brankov}\ \emph {et~al.}(2000)\citenamefont
  {Brankov}, \citenamefont {Danchev},\ and\ \citenamefont
  {Tonchev}}]{Brankov:book}%
  \BibitemOpen
  \bibfield  {author} {\bibinfo {author} {\bibfnamefont {J.~G.}\ \bibnamefont
  {Brankov}}, \bibinfo {author} {\bibfnamefont {D.~M.}\ \bibnamefont
  {Danchev}}, \ and\ \bibinfo {author} {\bibfnamefont {N.~S.}\ \bibnamefont
  {Tonchev}},\ }\href@noop {} {\emph {\bibinfo {title} {Theory of critical
  phenomena in finite size systems}}}\ (\bibinfo  {publisher} {World
  Scientific, Singapore},\ \bibinfo {year} {2000})\BibitemShut {NoStop}%
\bibitem [{\citenamefont {Gambassi}(2009)}]{Gambassi:2009conf}%
  \BibitemOpen
  \bibfield  {author} {\bibinfo {author} {\bibfnamefont {A.}~\bibnamefont
  {Gambassi}},\ }\href@noop {} {\bibfield  {journal} {\bibinfo  {journal} {J.
  Phys.: Conf. Ser.}\ }\textbf {\bibinfo {Volume} {161}},\ \bibinfo {pages}
  {012037} (\bibinfo {year} {2009})}\BibitemShut {NoStop}%
\bibitem [{\citenamefont {Garcia}\ and\ \citenamefont
  {Chan}(1999)}]{Garcia:1999}%
  \BibitemOpen
  \bibfield  {author} {\bibinfo {author} {\bibfnamefont {R.}~\bibnamefont
  {Garcia}}\ and\ \bibinfo {author} {\bibfnamefont {M.~H.~W.}\ \bibnamefont
  {Chan}},\ }\href {\doibase 10.1103/PhysRevLett.83.1187} {\bibfield  {journal}
  {\bibinfo  {journal} {Phys. Rev. Lett.}\ }\textbf {\bibinfo {Volume} {83}},\
  \bibinfo {pages} {1187} (\bibinfo {year} {1999})}\BibitemShut {NoStop}%
\bibitem [{\citenamefont {Garcia}\ and\ \citenamefont
  {Chan}(2002)}]{Garcia:2002}%
  \BibitemOpen
  \bibfield  {author} {\bibinfo {author} {\bibfnamefont {R.}~\bibnamefont
  {Garcia}}\ and\ \bibinfo {author} {\bibfnamefont {M.~H.~W.}\ \bibnamefont
  {Chan}},\ }\href@noop {} {\bibfield  {journal} {\bibinfo  {journal} {Phys.
  Rev. Lett.}\ }\textbf {\bibinfo {Volume} {88}},\ \bibinfo {pages} {086101}
  (\bibinfo {year} {2002})}\BibitemShut {NoStop}%
\bibitem [{\citenamefont {Fukuto}\ \emph {et~al.}(2005)\citenamefont {Fukuto},
  \citenamefont {Yano},\ and\ \citenamefont {Pershan}}]{Fukuto:2005}%
  \BibitemOpen
  \bibfield  {author} {\bibinfo {author} {\bibfnamefont {M.}~\bibnamefont
  {Fukuto}}, \bibinfo {author} {\bibfnamefont {Y.~F.}\ \bibnamefont {Yano}}, \
  and\ \bibinfo {author} {\bibfnamefont {P.~S.}\ \bibnamefont {Pershan}},\
  }\href {\doibase 10.1103/PhysRevLett.94.135702} {\bibfield  {journal}
  {\bibinfo  {journal} {Phys. Rev. Lett.}\ }\textbf {\bibinfo {Volume} {94}},\
  \bibinfo {pages} {135702} (\bibinfo {year} {2005})}\BibitemShut {NoStop}%
\bibitem [{\citenamefont {Ganshin}\ \emph {et~al.}(2006)\citenamefont
  {Ganshin}, \citenamefont {Scheidemantel}, \citenamefont {Garcia},\ and\
  \citenamefont {Chan}}]{Ganshin:2006}%
  \BibitemOpen
  \bibfield  {author} {\bibinfo {author} {\bibfnamefont {A.}~\bibnamefont
  {Ganshin}}, \bibinfo {author} {\bibfnamefont {S.}~\bibnamefont
  {Scheidemantel}}, \bibinfo {author} {\bibfnamefont {R.}~\bibnamefont
  {Garcia}}, \ and\ \bibinfo {author} {\bibfnamefont {M.~H.~W.}\ \bibnamefont
  {Chan}},\ }\href {\doibase 10.1103/PhysRevLett.97.075301} {\bibfield
  {journal} {\bibinfo  {journal} {Phys. Rev. Lett.}\ }\textbf {\bibinfo
  {Volume} {97}},\ \bibinfo {pages} {075301} (\bibinfo {year}
  {2006})}\BibitemShut {NoStop}%
\bibitem [{\citenamefont {Rafai}\ \emph {et~al.}(2007)\citenamefont {Rafai},
  \citenamefont {Bonn},\ and\ \citenamefont {Meunier}}]{Rafai:2007}%
  \BibitemOpen
  \bibfield  {author} {\bibinfo {author} {\bibfnamefont {S.}~\bibnamefont
  {Rafai}}, \bibinfo {author} {\bibfnamefont {D.}~\bibnamefont {Bonn}}, \ and\
  \bibinfo {author} {\bibfnamefont {J.}~\bibnamefont {Meunier}},\ }\href
  {\doibase 10.1016/j.physa.2007.07.072} {\bibfield  {journal} {\bibinfo
  {journal} {Physica A}\ }\textbf {\bibinfo {Volume} {386}},\ \bibinfo {pages}
  {31} (\bibinfo {year} {2007})}\BibitemShut {NoStop}%
\bibitem [{\citenamefont {Hertlein}\ \emph {et~al.}(2008)\citenamefont
  {Hertlein}, \citenamefont {Helden}, \citenamefont {Gambassi}, \citenamefont
  {Dietrich},\ and\ \citenamefont {Bechinger}}]{Hertlein:2008}%
  \BibitemOpen
  \bibfield  {author} {\bibinfo {author} {\bibfnamefont {C.}~\bibnamefont
  {Hertlein}}, \bibinfo {author} {\bibfnamefont {L.}~\bibnamefont {Helden}},
  \bibinfo {author} {\bibfnamefont {A.}~\bibnamefont {Gambassi}}, \bibinfo
  {author} {\bibfnamefont {S.}~\bibnamefont {Dietrich}}, \ and\ \bibinfo
  {author} {\bibfnamefont {C.}~\bibnamefont {Bechinger}},\ }\href {\doibase
  http://dx.doi.org/10.1038/nature06443} {\bibfield  {journal} {\bibinfo
  {journal} {Nature}\ }\textbf {\bibinfo {Volume} {451}},\ \bibinfo {pages}
  {172} (\bibinfo {year} {2008})}\BibitemShut {NoStop}%
\bibitem [{\citenamefont {Gambassi}\ \emph {et~al.}(2009)\citenamefont
  {Gambassi}, \citenamefont {Macio{\l}ek}, \citenamefont {Hertlein},
  \citenamefont {Nellen}, \citenamefont {Helden}, \citenamefont {Bechinger},\
  and\ \citenamefont {Dietrich}}]{Gambassi:2009}%
  \BibitemOpen
  \bibfield  {author} {\bibinfo {author} {\bibfnamefont {A.}~\bibnamefont
  {Gambassi}}, \bibinfo {author} {\bibfnamefont {A.}~\bibnamefont
  {Macio{\l}ek}}, \bibinfo {author} {\bibfnamefont {C.}~\bibnamefont
  {Hertlein}}, \bibinfo {author} {\bibfnamefont {U.}~\bibnamefont {Nellen}},
  \bibinfo {author} {\bibfnamefont {L.}~\bibnamefont {Helden}}, \bibinfo
  {author} {\bibfnamefont {C.}~\bibnamefont {Bechinger}}, \ and\ \bibinfo
  {author} {\bibfnamefont {S.}~\bibnamefont {Dietrich}},\ }\href@noop {}
  {\bibfield  {journal} {\bibinfo  {journal} {Phys. Rev. E}\ }\textbf {\bibinfo
  {Volume} {80}},\ \bibinfo {pages} {061143} (\bibinfo {year}
  {2009})}\BibitemShut {NoStop}%
\bibitem [{\citenamefont {Nellen}\ \emph {et~al.}(2009)\citenamefont {Nellen},
  \citenamefont {Helden},\ and\ \citenamefont {Bechinger}}]{Nellen:2009}%
  \BibitemOpen
  \bibfield  {author} {\bibinfo {author} {\bibfnamefont {U.}~\bibnamefont
  {Nellen}}, \bibinfo {author} {\bibfnamefont {L.}~\bibnamefont {Helden}}, \
  and\ \bibinfo {author} {\bibfnamefont {C.}~\bibnamefont {Bechinger}},\ }\href
  {\doibase 10.1209/0295-5075/88/26001} {\bibfield  {journal} {\bibinfo
  {journal} {EPL}\ }\textbf {\bibinfo {Volume} {88}},\ \bibinfo {pages} {26001}
  (\bibinfo {year} {2009})}\BibitemShut {NoStop}%
\bibitem [{\citenamefont {Binder}(1983)}]{Binder:1983}%
  \BibitemOpen
  \bibfield  {author} {\bibinfo {author} {\bibfnamefont {K.}~\bibnamefont
  {Binder}},\ }in\ \href@noop {} {\emph {\bibinfo {booktitle} {Phase
  Transitions and Critical Phenomena}}},\ Vol.~\bibinfo {Volume} {8},\ \bibinfo
  {editor} {edited by\ \bibinfo {editor} {\bibfnamefont {C.}~\bibnamefont
  {Domb}}\ and\ \bibinfo {editor} {\bibfnamefont {J.~L.}\ \bibnamefont
  {Lebowitz}}}\ (\bibinfo  {publisher} {Academic, London},\ \bibinfo {year}
  {1983})\ p.~\bibinfo {pages} {1}\BibitemShut {NoStop}%
\bibitem [{\citenamefont {Diehl}(1986)}]{Diehl:1986}%
  \BibitemOpen
  \bibfield  {author} {\bibinfo {author} {\bibfnamefont {H.~W.}\ \bibnamefont
  {Diehl}},\ }in\ \href@noop {} {\emph {\bibinfo {booktitle} {Phase Transitions
  and Critical Phenomena}}},\ Vol.~\bibinfo {Volume} {10},\ \bibinfo {editor}
  {edited by\ \bibinfo {editor} {\bibfnamefont {C.}~\bibnamefont {Domb}}\ and\
  \bibinfo {editor} {\bibfnamefont {J.~L.}\ \bibnamefont {Lebowitz}}}\
  (\bibinfo  {publisher} {Academic, London},\ \bibinfo {year} {1986})\
  p.~\bibinfo {pages} {75}\BibitemShut {NoStop}%
\bibitem [{\citenamefont {Bray}\ and\ \citenamefont {Moore}(1977)}]{bray:1977}%
  \BibitemOpen
  \bibfield  {author} {\bibinfo {author} {\bibfnamefont {A.~J.}\ \bibnamefont
  {Bray}}\ and\ \bibinfo {author} {\bibfnamefont {M.~A.}\ \bibnamefont
  {Moore}},\ }\href@noop {} {\bibfield  {journal} {\bibinfo  {journal} {J.
  Phys. A}\ }\textbf {\bibinfo {Volume} {10}},\ \bibinfo {pages} {1927}
  (\bibinfo {year} {1977})}\BibitemShut {NoStop}%
\bibitem [{\citenamefont {Burkhardt}\ and\ \citenamefont
  {Diehl}(1994)}]{burkhardt:1994}%
  \BibitemOpen
  \bibfield  {author} {\bibinfo {author} {\bibfnamefont {T.~W.}\ \bibnamefont
  {Burkhardt}}\ and\ \bibinfo {author} {\bibfnamefont {H.~W.}\ \bibnamefont
  {Diehl}},\ }\href@noop {} {\bibfield  {journal} {\bibinfo  {journal} {Phys.
  Rev. B}\ }\textbf {\bibinfo {Volume} {50}},\ \bibinfo {pages} {3894}
  (\bibinfo {year} {1994})}\BibitemShut {NoStop}%
\bibitem [{\citenamefont {Xia}\ and\ \citenamefont
  {Whitesides}(1998)}]{Xia:1998}%
  \BibitemOpen
  \bibfield  {author} {\bibinfo {author} {\bibfnamefont {Y.~N.}\ \bibnamefont
  {Xia}}\ and\ \bibinfo {author} {\bibfnamefont {G.~M.}\ \bibnamefont
  {Whitesides}},\ }\href@noop {} {\bibfield  {journal} {\bibinfo  {journal}
  {Annu. Rev. Mater. Sci.}\ }\textbf {\bibinfo {Volume} {28}},\ \bibinfo
  {pages} {153} (\bibinfo {year} {1998})}\BibitemShut {NoStop}%
\bibitem [{\citenamefont {Wang}\ \emph {et~al.}(1998)\citenamefont {Wang},
  \citenamefont {Thompson},\ and\ \citenamefont {Simmons}}]{Wang:1998}%
  \BibitemOpen
  \bibfield  {author} {\bibinfo {author} {\bibfnamefont {J.}~\bibnamefont
  {Wang}}, \bibinfo {author} {\bibfnamefont {D.~K.}\ \bibnamefont {Thompson}},
  \ and\ \bibinfo {author} {\bibfnamefont {J.~G.}\ \bibnamefont {Simmons}},\
  }\href@noop {} {\bibfield  {journal} {\bibinfo  {journal} {J. Electrochem.
  Soc.}\ }\textbf {\bibinfo {Volume} {145}},\ \bibinfo {pages} {2931} (\bibinfo
  {year} {1998})}\BibitemShut {NoStop}%
\bibitem [{\citenamefont {Tolfree}(1998)}]{Tolfree:1998}%
  \BibitemOpen
  \bibfield  {author} {\bibinfo {author} {\bibfnamefont {D.~W.~L.}\
  \bibnamefont {Tolfree}},\ }\href@noop {} {\bibfield  {journal} {\bibinfo
  {journal} {Rep. Prog. Phys.}\ }\textbf {\bibinfo {Volume} {61}},\ \bibinfo
  {pages} {313} (\bibinfo {year} {1998})}\BibitemShut {NoStop}%
\bibitem [{\citenamefont {Nie}\ and\ \citenamefont
  {Kumacheva}(2008)}]{Nie:2008}%
  \BibitemOpen
  \bibfield  {author} {\bibinfo {author} {\bibfnamefont {Z.~H.}\ \bibnamefont
  {Nie}}\ and\ \bibinfo {author} {\bibfnamefont {E.}~\bibnamefont
  {Kumacheva}},\ }\href@noop {} {\bibfield  {journal} {\bibinfo  {journal}
  {Nature Mater.}\ }\textbf {\bibinfo {Volume} {7}},\ \bibinfo {pages} {277}
  (\bibinfo {year} {2008})}\BibitemShut {NoStop}%
\bibitem [{\citenamefont {Thorsen}\ \emph {et~al.}(2002)\citenamefont
  {Thorsen}, \citenamefont {Maerkl},\ and\ \citenamefont
  {Quake}}]{Thorsen:2002}%
  \BibitemOpen
  \bibfield  {author} {\bibinfo {author} {\bibfnamefont {T.}~\bibnamefont
  {Thorsen}}, \bibinfo {author} {\bibfnamefont {S.~J.}\ \bibnamefont {Maerkl}},
  \ and\ \bibinfo {author} {\bibfnamefont {S.~R.}\ \bibnamefont {Quake}},\
  }\href@noop {} {\bibfield  {journal} {\bibinfo  {journal} {Science}\ }\textbf
  {\bibinfo {Volume} {298}},\ \bibinfo {pages} {580} (\bibinfo {year}
  {2002})}\BibitemShut {NoStop}%
\bibitem [{\citenamefont {Sprenger}\ \emph {et~al.}(2006)\citenamefont
  {Sprenger}, \citenamefont {Schlesener},\ and\ \citenamefont
  {Dietrich}}]{Sprenger:2006}%
  \BibitemOpen
  \bibfield  {author} {\bibinfo {author} {\bibfnamefont {M.}~\bibnamefont
  {Sprenger}}, \bibinfo {author} {\bibfnamefont {F.}~\bibnamefont
  {Schlesener}}, \ and\ \bibinfo {author} {\bibfnamefont {S.}~\bibnamefont
  {Dietrich}},\ }\href@noop {} {\bibfield  {journal} {\bibinfo  {journal} {J.
  Chem. Phys.}\ }\textbf {\bibinfo {Volume} {124}},\ \bibinfo {pages} {134703}
  (\bibinfo {year} {2006})}\BibitemShut {NoStop}%
\bibitem [{\citenamefont {Tr{\"o}ndle}\ \emph {et~al.}(2009)\citenamefont
  {Tr{\"o}ndle}, \citenamefont {Kondrat}, \citenamefont {Gambassi},
  \citenamefont {Harnau},\ and\ \citenamefont {Dietrich}}]{Troendle:2009}%
  \BibitemOpen
  \bibfield  {author} {\bibinfo {author} {\bibfnamefont {M.}~\bibnamefont
  {Tr{\"o}ndle}}, \bibinfo {author} {\bibfnamefont {S.}~\bibnamefont
  {Kondrat}}, \bibinfo {author} {\bibfnamefont {A.}~\bibnamefont {Gambassi}},
  \bibinfo {author} {\bibfnamefont {L.}~\bibnamefont {Harnau}}, \ and\ \bibinfo
  {author} {\bibfnamefont {S.}~\bibnamefont {Dietrich}},\ }\href {\doibase
  10.1209/0295-5075/88/40004} {\bibfield  {journal} {\bibinfo  {journal} {EPL}\
  }\textbf {\bibinfo {Volume} {88}},\ \bibinfo {pages} {40004} (\bibinfo {year}
  {2009})}\BibitemShut {NoStop}%
\bibitem [{\citenamefont {Tr{\"o}ndle}\ \emph {et~al.}(2010)\citenamefont
  {Tr{\"o}ndle}, \citenamefont {Kondrat}, \citenamefont {Gambassi},
  \citenamefont {Harnau},\ and\ \citenamefont {Dietrich}}]{Troendle:2010}%
  \BibitemOpen
  \bibfield  {author} {\bibinfo {author} {\bibfnamefont {M.}~\bibnamefont
  {Tr{\"o}ndle}}, \bibinfo {author} {\bibfnamefont {S.}~\bibnamefont
  {Kondrat}}, \bibinfo {author} {\bibfnamefont {A.}~\bibnamefont {Gambassi}},
  \bibinfo {author} {\bibfnamefont {L.}~\bibnamefont {Harnau}}, \ and\ \bibinfo
  {author} {\bibfnamefont {S.}~\bibnamefont {Dietrich}},\ }\href {\doibase
  10.1063/1.3464770} {\bibfield  {journal} {\bibinfo  {journal} {J. Chem.
  Phys.}\ }\textbf {\bibinfo {Volume} {133}},\ \bibinfo {pages} {074702}
  (\bibinfo {year} {2010})}\BibitemShut {NoStop}%
\bibitem [{\citenamefont {Gambassi}\ and\ \citenamefont
  {Dietrich}(2011)}]{Gambassi:2011}%
  \BibitemOpen
  \bibfield  {author} {\bibinfo {author} {\bibfnamefont {A.}~\bibnamefont
  {Gambassi}}\ and\ \bibinfo {author} {\bibfnamefont {S.}~\bibnamefont
  {Dietrich}},\ }\href {\doibase 10.1039/C0SM00635A} {\bibfield  {journal}
  {\bibinfo  {journal} {Soft Matter}\ }\textbf {\bibinfo {Volume} {7}},\
  \bibinfo {pages} {1247} (\bibinfo {year} {2011})}\BibitemShut {NoStop}%
\bibitem [{\citenamefont {Parisen~Toldin}\ and\ \citenamefont
  {Dietrich}(2010)}]{Parisen:2010}%
  \BibitemOpen
  \bibfield  {author} {\bibinfo {author} {\bibfnamefont {F.}~\bibnamefont
  {Parisen~Toldin}}\ and\ \bibinfo {author} {\bibfnamefont {S.}~\bibnamefont
  {Dietrich}},\ }\href {http://stacks.iop.org/1742-5468/2010/i=11/a=P11003}
  {\bibfield  {journal} {\bibinfo  {journal} {J. Stat. Mech.}\ ,\ \bibinfo
  {pages} {P11003}} (\bibinfo {year} {2010})}\BibitemShut {NoStop}%
\bibitem [{\citenamefont {Parisen~Toldin}\ \emph {et~al.}(2013)\citenamefont
  {Parisen~Toldin}, \citenamefont {Tr\"{o}ndle},\ and\ \citenamefont
  {Dietrich}}]{Parisen:2013}%
  \BibitemOpen
  \bibfield  {author} {\bibinfo {author} {\bibfnamefont {F.}~\bibnamefont
  {Parisen~Toldin}}, \bibinfo {author} {\bibfnamefont {M.}~\bibnamefont
  {Tr\"{o}ndle}}, \ and\ \bibinfo {author} {\bibfnamefont {S.}~\bibnamefont
  {Dietrich}},\ }\href@noop {} {\bibfield  {journal} {\bibinfo  {journal}
  {Phys. Rev. E}\ }\textbf {\bibinfo {Volume} {88}},\ \bibinfo {pages} {052110}
  (\bibinfo {year} {2013})}\BibitemShut {NoStop}%
\bibitem [{\citenamefont {Parisen~Toldin}\ \emph {et~al.}(2014)\citenamefont
  {Parisen~Toldin}, \citenamefont {Tr\"{o}ndle},\ and\ \citenamefont
  {Dietrich}}]{Parisen:2014}%
  \BibitemOpen
  \bibfield  {author} {\bibinfo {author} {\bibfnamefont {F.}~\bibnamefont
  {Parisen~Toldin}}, \bibinfo {author} {\bibfnamefont {M.}~\bibnamefont
  {Tr\"{o}ndle}}, \ and\ \bibinfo {author} {\bibfnamefont {S.}~\bibnamefont
  {Dietrich}},\ }\href@noop {} {\bibfield  {journal} {\bibinfo  {journal} in preparation}  (\bibinfo {year} {2014}) }
  \BibitemShut 
  {NoStop}%
\bibitem [{\citenamefont {Labb{\'e}-Laurent}\ \emph {et~al.}(2014)\citenamefont
  {Labb{\'e}-Laurent}, \citenamefont {Tr{\"o}ndle}, \citenamefont {Harnau},\
  and\ \citenamefont {Dietrich}}]{Labbe:2014}%
  \BibitemOpen
  \bibfield  {author} {\bibinfo {author} {\bibfnamefont {M.}~\bibnamefont
  {Labb{\'e}-Laurent}}, \bibinfo {author} {\bibfnamefont {M.}~\bibnamefont
  {Tr{\"o}ndle}}, \bibinfo {author} {\bibfnamefont {L.}~\bibnamefont {Harnau}},
  \ and\ \bibinfo {author} {\bibfnamefont {S.}~\bibnamefont {Dietrich}},\
  }\href {\doibase 10.1039/C3SM52858H} {\bibfield  {journal} {\bibinfo
  {journal} {Soft Matter}\ }\textbf {\bibinfo {Volume} {10}},\ \bibinfo {pages}
  {2270} (\bibinfo {year} {2014})}\BibitemShut {NoStop}%
\bibitem [{\citenamefont {Soyka}\ \emph {et~al.}(2008)\citenamefont {Soyka},
  \citenamefont {Zvyagolskaya}, \citenamefont {Hertlein}, \citenamefont
  {Helden},\ and\ \citenamefont {Bechinger}}]{Soyka:2008}%
  \BibitemOpen
  \bibfield  {author} {\bibinfo {author} {\bibfnamefont {F.}~\bibnamefont
  {Soyka}}, \bibinfo {author} {\bibfnamefont {O.}~\bibnamefont {Zvyagolskaya}},
  \bibinfo {author} {\bibfnamefont {C.}~\bibnamefont {Hertlein}}, \bibinfo
  {author} {\bibfnamefont {L.}~\bibnamefont {Helden}}, \ and\ \bibinfo {author}
  {\bibfnamefont {C.}~\bibnamefont {Bechinger}},\ }\href {\doibase
  10.1103/PhysRevLett.101.208301} {\bibfield  {journal} {\bibinfo  {journal}
  {Phys. Rev. Lett.}\ }\textbf {\bibinfo {Volume} {101}},\ \bibinfo {pages}
  {208301} (\bibinfo {year} {2008})}\BibitemShut {NoStop}%
\bibitem [{\citenamefont {Tr{\"o}ndle}\ \emph {et~al.}(2011)\citenamefont
  {Tr{\"o}ndle}, \citenamefont {Zvyagolskaya}, \citenamefont {Gambassi},
  \citenamefont {Vogt}, \citenamefont {Harnau}, \citenamefont {Bechinger},\
  and\ \citenamefont {Dietrich}}]{Troendle:2011}%
  \BibitemOpen
  \bibfield  {author} {\bibinfo {author} {\bibfnamefont {M.}~\bibnamefont
  {Tr{\"o}ndle}}, \bibinfo {author} {\bibfnamefont {O.}~\bibnamefont
  {Zvyagolskaya}}, \bibinfo {author} {\bibfnamefont {A.}~\bibnamefont
  {Gambassi}}, \bibinfo {author} {\bibfnamefont {D.}~\bibnamefont {Vogt}},
  \bibinfo {author} {\bibfnamefont {L.}~\bibnamefont {Harnau}}, \bibinfo
  {author} {\bibfnamefont {C.}~\bibnamefont {Bechinger}}, \ and\ \bibinfo
  {author} {\bibfnamefont {S.}~\bibnamefont {Dietrich}},\ }\href
  {http://www.informaworld.com/10.1080/00268976.2011.553639} {\bibfield
  {journal} {\bibinfo  {journal} {Mol. Phys.}\ }\textbf {\bibinfo {Volume}
  {109}},\ \bibinfo {pages} {1169} (\bibinfo {year} {2011})}\BibitemShut
  {NoStop}%
\bibitem [{\citenamefont {Emig}\ \emph {et~al.}(2001)\citenamefont {Emig},
  \citenamefont {Hanke}, \citenamefont {Golestanian},\ and\ \citenamefont
  {Kardar}}]{Emig:2001}%
  \BibitemOpen
  \bibfield  {author} {\bibinfo {author} {\bibfnamefont {T.}~\bibnamefont
  {Emig}}, \bibinfo {author} {\bibfnamefont {A.}~\bibnamefont {Hanke}},
  \bibinfo {author} {\bibfnamefont {R.}~\bibnamefont {Golestanian}}, \ and\
  \bibinfo {author} {\bibfnamefont {M.}~\bibnamefont {Kardar}},\ }\href@noop {}
  {\bibfield  {journal} {\bibinfo  {journal} {Phys. Rev. Lett.}\ }\textbf
  {\bibinfo {Volume} {87}},\ \bibinfo {pages} {260402} (\bibinfo {year}
  {2001})}\BibitemShut {NoStop}%
\bibitem [{\citenamefont {Emig}\ \emph {et~al.}(2003)\citenamefont {Emig},
  \citenamefont {Hanke}, \citenamefont {Golestanian},\ and\ \citenamefont
  {Kardar}}]{Emig:2003}%
  \BibitemOpen
  \bibfield  {author} {\bibinfo {author} {\bibfnamefont {T.}~\bibnamefont
  {Emig}}, \bibinfo {author} {\bibfnamefont {A.}~\bibnamefont {Hanke}},
  \bibinfo {author} {\bibfnamefont {R.}~\bibnamefont {Golestanian}}, \ and\
  \bibinfo {author} {\bibfnamefont {M.}~\bibnamefont {Kardar}},\ }\href@noop {}
  {\bibfield  {journal} {\bibinfo  {journal} {Phys. Rev. A}\ }\textbf {\bibinfo
  {Volume} {67}},\ \bibinfo {pages} {022114} (\bibinfo {year}
  {2003})}\BibitemShut {NoStop}%
\bibitem [{\citenamefont {van Zwol}\ \emph {et~al.}(2008)\citenamefont {van
  Zwol}, \citenamefont {Palasantzas},\ and\ \citenamefont
  {de~hosson}}]{Zwol:2008}%
  \BibitemOpen
  \bibfield  {author} {\bibinfo {author} {\bibfnamefont {P.~J.}\ \bibnamefont
  {van Zwol}}, \bibinfo {author} {\bibfnamefont {G.}~\bibnamefont
  {Palasantzas}}, \ and\ \bibinfo {author} {\bibfnamefont {J.~T.~M.}\
  \bibnamefont {de~hosson}},\ }\href@noop {} {\bibfield  {journal} {\bibinfo
  {journal} {Appl. Phys. Lett.}\ }\textbf {\bibinfo {Volume} {92}} (\bibinfo
  {year} {2008})}\BibitemShut {NoStop}%
\bibitem [{\citenamefont {Lambrecht}\ and\ \citenamefont
  {Marachevsky}(2008)}]{Lambrecht:2008}%
  \BibitemOpen
  \bibfield  {author} {\bibinfo {author} {\bibfnamefont {A.}~\bibnamefont
  {Lambrecht}}\ and\ \bibinfo {author} {\bibfnamefont {V.~N.}\ \bibnamefont
  {Marachevsky}},\ }\href {\doibase 10.1103/PhysRevLett.101.160403} {\bibfield
  {journal} {\bibinfo  {journal} {Phys. Rev. Lett.}\ }\textbf {\bibinfo
  {Volume} {101}},\ \bibinfo {pages} {160403} (\bibinfo {year}
  {2008})}\BibitemShut {NoStop}%
\bibitem [{\citenamefont {Rodriguez}\ \emph {et~al.}(2011)\citenamefont
  {Rodriguez}, \citenamefont {Capasso},\ and\ \citenamefont
  {Johnson}}]{Rodriguez:2011}%
  \BibitemOpen
  \bibfield  {author} {\bibinfo {author} {\bibfnamefont {A.~W.}\ \bibnamefont
  {Rodriguez}}, \bibinfo {author} {\bibfnamefont {F.}~\bibnamefont {Capasso}},
  \ and\ \bibinfo {author} {\bibfnamefont {S.~G.}\ \bibnamefont {Johnson}},\
  }\href {http://dx.doi.org/10.1038/nphoton.2011.39} {\bibfield  {journal}
  {\bibinfo  {journal} {Nature Photon.}\ }\textbf {\bibinfo {Volume} {5}},\
  \bibinfo {pages} {211} (\bibinfo {year} {2011})}\BibitemShut {NoStop}%
\bibitem [{\citenamefont {Lussange}\ \emph {et~al.}(2012)\citenamefont
  {Lussange}, \citenamefont {Gu\'erout},\ and\ \citenamefont
  {Lambrecht}}]{Lussange:2012}%
  \BibitemOpen
  \bibfield  {author} {\bibinfo {author} {\bibfnamefont {J.}~\bibnamefont
  {Lussange}}, \bibinfo {author} {\bibfnamefont {R.}~\bibnamefont {Gu\'erout}},
  \ and\ \bibinfo {author} {\bibfnamefont {A.}~\bibnamefont {Lambrecht}},\
  }\href {\doibase 10.1103/PhysRevA.86.062502} {\bibfield  {journal} {\bibinfo
  {journal} {Phys. Rev. A}\ }\textbf {\bibinfo {Volume} {86}},\ \bibinfo
  {pages} {062502} (\bibinfo {year} {2012})}\BibitemShut {NoStop}%
\bibitem [{\citenamefont {Chen}\ \emph {et~al.}(2002)\citenamefont {Chen},
  \citenamefont {Mohideen}, \citenamefont {Klimchitskaya},\ and\ \citenamefont
  {Mostepanenko}}]{Chen:2002}%
  \BibitemOpen
  \bibfield  {author} {\bibinfo {author} {\bibfnamefont {F.}~\bibnamefont
  {Chen}}, \bibinfo {author} {\bibfnamefont {U.}~\bibnamefont {Mohideen}},
  \bibinfo {author} {\bibfnamefont {G.~L.}\ \bibnamefont {Klimchitskaya}}, \
  and\ \bibinfo {author} {\bibfnamefont {V.~M.}\ \bibnamefont {Mostepanenko}},\
  }\href {\doibase 10.1103/PhysRevLett.88.101801} {\bibfield  {journal}
  {\bibinfo  {journal} {Phys. Rev. Lett.}\ }\textbf {\bibinfo {Volume} {88}},\
  \bibinfo {pages} {101801} (\bibinfo {year} {2002})}\BibitemShut {NoStop}%
\bibitem [{\citenamefont {B{\"u}scher}\ and\ \citenamefont
  {Emig}(2005)}]{Emig:2005}%
  \BibitemOpen
  \bibfield  {author} {\bibinfo {author} {\bibfnamefont {R.}~\bibnamefont
  {B{\"u}scher}}\ and\ \bibinfo {author} {\bibfnamefont {T.}~\bibnamefont
  {Emig}},\ }\href@noop {} {\bibfield  {journal} {\bibinfo  {journal} {Phys.
  Rev. Lett.}\ }\textbf {\bibinfo {Volume} {94}},\ \bibinfo {pages} {133901}
  (\bibinfo {year} {2005})}\BibitemShut {NoStop}%
\bibitem [{\citenamefont {Rodrigues}\ \emph {et~al.}(2006)\citenamefont
  {Rodrigues}, \citenamefont {Neto}, \citenamefont {Lambrecht},\ and\
  \citenamefont {Reynaud}}]{Rodrigues:2006}%
  \BibitemOpen
  \bibfield  {author} {\bibinfo {author} {\bibfnamefont {R.~B.}\ \bibnamefont
  {Rodrigues}}, \bibinfo {author} {\bibfnamefont {P.~A.~M.}\ \bibnamefont
  {Neto}}, \bibinfo {author} {\bibfnamefont {A.}~\bibnamefont {Lambrecht}}, \
  and\ \bibinfo {author} {\bibfnamefont {S.}~\bibnamefont {Reynaud}},\
  }\href@noop {} {\bibfield  {journal} {\bibinfo  {journal} {Phys. Rev. Lett.}\
  }\textbf {\bibinfo {Volume} {96}},\ \bibinfo {pages} {100402} (\bibinfo
  {year} {2006})}\BibitemShut {NoStop}%
\bibitem [{\citenamefont {Dalvit}\ \emph {et~al.}(2008)\citenamefont {Dalvit},
  \citenamefont {Maia~Neto}, \citenamefont {A.},\ and\ \citenamefont
  {S.}}]{Dalvit:2008}%
  \BibitemOpen
  \bibfield  {author} {\bibinfo {author} {\bibfnamefont {D.~A.~R.}\
  \bibnamefont {Dalvit}}, \bibinfo {author} {\bibfnamefont {P.~A.}\
  \bibnamefont {Maia~Neto}}, \bibinfo {author} {\bibfnamefont {L.}~\bibnamefont
  {A.}}, \ and\ \bibinfo {author} {\bibfnamefont {R.}~\bibnamefont {S.}},\
  }\href@noop {} {\bibfield  {journal} {\bibinfo  {journal} {J. Phys. A}\
  }\textbf {\bibinfo {Volume} {41}},\ \bibinfo {pages} {164028} (\bibinfo
  {year} {2008})}\BibitemShut {NoStop}%
\bibitem [{\citenamefont {Ashourvan}\ \emph {et~al.}(2007)\citenamefont
  {Ashourvan}, \citenamefont {Miri},\ and\ \citenamefont
  {Golestanian}}]{Ashourvan:2007}%
  \BibitemOpen
  \bibfield  {author} {\bibinfo {author} {\bibfnamefont {A.}~\bibnamefont
  {Ashourvan}}, \bibinfo {author} {\bibfnamefont {M.}~\bibnamefont {Miri}}, \
  and\ \bibinfo {author} {\bibfnamefont {R.}~\bibnamefont {Golestanian}},\
  }\href@noop {} {\bibfield  {journal} {\bibinfo  {journal} {Phys. Rev. Lett.}\
  }\textbf {\bibinfo {Volume} {98}},\ \bibinfo {pages} {140801} (\bibinfo
  {year} {2007})}\BibitemShut {NoStop}%
\bibitem [{\citenamefont {Emig}(2007)}]{Emig:2007}%
  \BibitemOpen
  \bibfield  {author} {\bibinfo {author} {\bibfnamefont {T.}~\bibnamefont
  {Emig}},\ }\href@noop {} {\bibfield  {journal} {\bibinfo  {journal} {Phys.
  Rev. Lett.}\ }\textbf {\bibinfo {Volume} {98}},\ \bibinfo {pages} {160801}
  (\bibinfo {year} {2007})}\BibitemShut {NoStop}%
\bibitem [{\citenamefont {Miri}\ and\ \citenamefont
  {Golestanian}(2008)}]{Miri:2008}%
  \BibitemOpen
  \bibfield  {author} {\bibinfo {author} {\bibfnamefont {M.}~\bibnamefont
  {Miri}}\ and\ \bibinfo {author} {\bibfnamefont {R.}~\bibnamefont
  {Golestanian}},\ }\href@noop {} {\bibfield  {journal} {\bibinfo  {journal}
  {Appl. Phys. Lett.}\ }\textbf {\bibinfo {Volume} {92}} (\bibinfo {year}
  {2008})}\BibitemShut {NoStop}%
\bibitem [{\citenamefont {Rodrigues}\ \emph {et~al.}(2008)\citenamefont
  {Rodrigues}, \citenamefont {Maia}, \citenamefont {Lambrecht},\ and\
  \citenamefont {Reynaud}}]{Rodrigues:2008}%
  \BibitemOpen
  \bibfield  {author} {\bibinfo {author} {\bibfnamefont {R.~B.}\ \bibnamefont
  {Rodrigues}}, \bibinfo {author} {\bibfnamefont {P.~A.}\ \bibnamefont {Maia}},
  \bibinfo {author} {\bibfnamefont {A.}~\bibnamefont {Lambrecht}}, \ and\
  \bibinfo {author} {\bibfnamefont {S.}~\bibnamefont {Reynaud}},\ }\href@noop
  {} {\bibfield  {journal} {\bibinfo  {journal} {J. Phys. A}\ }\textbf
  {\bibinfo {Volume} {41}} (\bibinfo {year} {2008})}\BibitemShut {NoStop}%
\bibitem [{\citenamefont {Chiu}\ \emph {et~al.}(2010)\citenamefont {Chiu},
  \citenamefont {Klimchitskaya}, \citenamefont {Marachevsky}, \citenamefont
  {Mostepanenko},\ and\ \citenamefont {Mohideen}}]{Chiu:2010}%
  \BibitemOpen
  \bibfield  {author} {\bibinfo {author} {\bibfnamefont {H.-C.}\ \bibnamefont
  {Chiu}}, \bibinfo {author} {\bibfnamefont {G.~L.}\ \bibnamefont
  {Klimchitskaya}}, \bibinfo {author} {\bibfnamefont {V.~N.}\ \bibnamefont
  {Marachevsky}}, \bibinfo {author} {\bibfnamefont {V.~M.}\ \bibnamefont
  {Mostepanenko}}, \ and\ \bibinfo {author} {\bibfnamefont {U.}~\bibnamefont
  {Mohideen}},\ }\href {\doibase 10.1103/PhysRevB.81.115417} {\bibfield
  {journal} {\bibinfo  {journal} {Phys. Rev. B}\ }\textbf {\bibinfo {Volume}
  {81}},\ \bibinfo {pages} {115417} (\bibinfo {year} {2010})}\BibitemShut
  {NoStop}%
\bibitem [{\citenamefont {Chan}\ \emph {et~al.}(2008)\citenamefont {Chan},
  \citenamefont {Bao}, \citenamefont {Zou}, \citenamefont {Cirelli},
  \citenamefont {Klemens}, \citenamefont {Mansfield},\ and\ \citenamefont
  {Pai}}]{Chan:2008}%
  \BibitemOpen
  \bibfield  {author} {\bibinfo {author} {\bibfnamefont {H.~B.}\ \bibnamefont
  {Chan}}, \bibinfo {author} {\bibfnamefont {Y.}~\bibnamefont {Bao}}, \bibinfo
  {author} {\bibfnamefont {J.}~\bibnamefont {Zou}}, \bibinfo {author}
  {\bibfnamefont {R.~A.}\ \bibnamefont {Cirelli}}, \bibinfo {author}
  {\bibfnamefont {F.}~\bibnamefont {Klemens}}, \bibinfo {author} {\bibfnamefont
  {W.~M.}\ \bibnamefont {Mansfield}}, \ and\ \bibinfo {author} {\bibfnamefont
  {C.~S.}\ \bibnamefont {Pai}},\ }\href@noop {} {\bibfield  {journal} {\bibinfo
   {journal} {Phys. Rev. Lett.}\ }\textbf {\bibinfo {Volume} {101}},\ \bibinfo
   {pages} {030401} (\bibinfo {year} {2008})}{; \textit{ibid.}
  \textbf{107}, 019901(E) (2011)}\BibitemShut {NoStop}%
\bibitem [{\citenamefont {Bao}\ \emph {et~al.}(2010)\citenamefont {Bao},
  \citenamefont {Gu\'erout}, \citenamefont {Lussange}, \citenamefont
  {Lambrecht}, \citenamefont {Cirelli}, \citenamefont {Klemens}, \citenamefont
  {Mansfield}, \citenamefont {Pai},\ and\ \citenamefont {Chan}}]{Bao:2010}%
  \BibitemOpen
  \bibfield  {author} {\bibinfo {author} {\bibfnamefont {Y.}~\bibnamefont
  {Bao}}, \bibinfo {author} {\bibfnamefont {R.}~\bibnamefont {Gu\'erout}},
  \bibinfo {author} {\bibfnamefont {J.}~\bibnamefont {Lussange}}, \bibinfo
  {author} {\bibfnamefont {A.}~\bibnamefont {Lambrecht}}, \bibinfo {author}
  {\bibfnamefont {R.~A.}\ \bibnamefont {Cirelli}}, \bibinfo {author}
  {\bibfnamefont {F.}~\bibnamefont {Klemens}}, \bibinfo {author} {\bibfnamefont
  {W.~M.}\ \bibnamefont {Mansfield}}, \bibinfo {author} {\bibfnamefont {C.~S.}\
  \bibnamefont {Pai}}, \ and\ \bibinfo {author} {\bibfnamefont {H.~B.}\
  \bibnamefont {Chan}},\ }\href {\doibase 10.1103/PhysRevLett.105.250402}
  {\bibfield  {journal} {\bibinfo  {journal} {Phys. Rev. Lett.}\ }\textbf
  {\bibinfo {Volume} {105}},\ \bibinfo {pages} {250402} (\bibinfo {year}
  {2010})}\BibitemShut {NoStop}%
\bibitem [{\citenamefont {Gu\'erout}\ \emph {et~al.}(2013)\citenamefont
  {Gu\'erout}, \citenamefont {Lussange}, \citenamefont {Chan}, \citenamefont
  {Lambrecht},\ and\ \citenamefont {Reynaud}}]{Guerout:2013}%
  \BibitemOpen
  \bibfield  {author} {\bibinfo {author} {\bibfnamefont {R.}~\bibnamefont
  {Gu\'erout}}, \bibinfo {author} {\bibfnamefont {J.}~\bibnamefont {Lussange}},
  \bibinfo {author} {\bibfnamefont {H.~B.}\ \bibnamefont {Chan}}, \bibinfo
  {author} {\bibfnamefont {A.}~\bibnamefont {Lambrecht}}, \ and\ \bibinfo
  {author} {\bibfnamefont {S.}~\bibnamefont {Reynaud}},\ }\href {\doibase
  10.1103/PhysRevA.87.052514} {\bibfield  {journal} {\bibinfo  {journal} {Phys.
  Rev. A}\ }\textbf {\bibinfo {Volume} {87}},\ \bibinfo {pages} {052514}
  (\bibinfo {year} {2013})}\BibitemShut {NoStop}%
\bibitem [{\citenamefont {Intravaia}\ \emph {et~al.}(2013)\citenamefont
  {Intravaia}, \citenamefont {Koev}, \citenamefont {Jung}, \citenamefont
  {Talin}, \citenamefont {Davids}, \citenamefont {Decca}, \citenamefont
  {Aksyuk}, \citenamefont {Dalvit},\ and\ \citenamefont
  {LÃ³pez}}]{Intravaia:2013}%
  \BibitemOpen
  \bibfield  {author} {\bibinfo {author} {\bibfnamefont {F.}~\bibnamefont
  {Intravaia}}, \bibinfo {author} {\bibfnamefont {S.}~\bibnamefont {Koev}},
  \bibinfo {author} {\bibfnamefont {I.~W.}\ \bibnamefont {Jung}}, \bibinfo
  {author} {\bibfnamefont {A.~A.}\ \bibnamefont {Talin}}, \bibinfo {author}
  {\bibfnamefont {P.~S.}\ \bibnamefont {Davids}}, \bibinfo {author}
  {\bibfnamefont {R.~S.}\ \bibnamefont {Decca}}, \bibinfo {author}
  {\bibfnamefont {V.~A.}\ \bibnamefont {Aksyuk}}, \bibinfo {author}
  {\bibfnamefont {D.~A.~R.}\ \bibnamefont {Dalvit}}, \ and\ \bibinfo {author}
  {\bibfnamefont {D.}~\bibnamefont {Lopez}},\ }\href
  {http://dx.doi.org/10.1038/ncomms3515} {\bibfield  {journal} {\bibinfo
  {journal} {Nature Commun.}\ }\textbf {\bibinfo {Volume} {4}},\  (\bibinfo {year}
  {2013})}\BibitemShut {NoStop}%
\bibitem [{\citenamefont {Cardy}(1983)}]{Cardy:1983}%
  \BibitemOpen
  \bibfield  {author} {\bibinfo {author} {\bibfnamefont {J.~L.}\ \bibnamefont
  {Cardy}},\ }\href@noop {} {\bibfield  {journal} {\bibinfo  {journal} {J.
  Phys. A}\ }\textbf {\bibinfo {Volume} {16}},\ \bibinfo {pages} {3617}
  (\bibinfo {year} {1983})}\BibitemShut {NoStop}%
\bibitem [{\citenamefont {Hanke}\ \emph {et~al.}(1999)\citenamefont {Hanke},
  \citenamefont {Krech}, \citenamefont {Schlesener},\ and\ \citenamefont
  {Dietrich}}]{Hanke:1999}%
  \BibitemOpen
  \bibfield  {author} {\bibinfo {author} {\bibfnamefont {A.}~\bibnamefont
  {Hanke}}, \bibinfo {author} {\bibfnamefont {M.}~\bibnamefont {Krech}},
  \bibinfo {author} {\bibfnamefont {F.}~\bibnamefont {Schlesener}}, \ and\
  \bibinfo {author} {\bibfnamefont {S.}~\bibnamefont {Dietrich}},\ }\href@noop
  {} {\bibfield  {journal} {\bibinfo  {journal} {Phys. Rev. E}\ }\textbf
  {\bibinfo {Volume} {60}},\ \bibinfo {pages} {5163} (\bibinfo {year}
  {1999})}\BibitemShut {NoStop}%
\bibitem [{\citenamefont {Pal{\'{a}}gyi}\ and\ \citenamefont
  {Dietrich}(2004)}]{Palagyi:2004}%
  \BibitemOpen
  \bibfield  {author} {\bibinfo {author} {\bibfnamefont {G.}~\bibnamefont
  {Pal{\'{a}}gyi}}\ and\ \bibinfo {author} {\bibfnamefont {S.}~\bibnamefont
  {Dietrich}},\ }\href@noop {} {\bibfield  {journal} {\bibinfo  {journal}
  {Phys. Rev. E}\ }\textbf {\bibinfo {Volume} {70}},\ \bibinfo {pages} {046144}
  (\bibinfo {year} {2004})}\BibitemShut {NoStop}%
\bibitem [{\citenamefont {Tr{\"o}ndle}\ \emph {et~al.}(2008)\citenamefont
  {Tr{\"o}ndle}, \citenamefont {Harnau},\ and\ \citenamefont
  {Dietrich}}]{Troendle:2008}%
  \BibitemOpen
  \bibfield  {author} {\bibinfo {author} {\bibfnamefont {M.}~\bibnamefont
  {Tr{\"o}ndle}}, \bibinfo {author} {\bibfnamefont {L.}~\bibnamefont {Harnau}},
  \ and\ \bibinfo {author} {\bibfnamefont {S.}~\bibnamefont {Dietrich}},\
  }\href {\doibase 10.1063/1.2977999} {\bibfield  {journal} {\bibinfo
  {journal} {J. Chem. Phys.}\ }\textbf {\bibinfo {Volume} {129}},\ \bibinfo
  {pages} {124716} (\bibinfo {year} {2008})}\BibitemShut {NoStop}%
\bibitem {Bimonte:2014}%
  \BibitemOpen
  \bibfield  {author} {\bibinfo {author} {\bibfnamefont {G.}~\bibnamefont
  {Bimonte}}, \bibinfo {author} {\bibfnamefont {T.}~\bibnamefont {Emig}},
  \ and\ \bibinfo {author} {\bibfnamefont {M.}~\bibnamefont {Kardar}}},\
  preprint arXiv:1406.0962 (2014)\BibitemShut {NoStop}%
\bibitem [{\citenamefont {Barber}(1983)}]{barber:1983}%
  \BibitemOpen
  \bibfield  {author} {\bibinfo {author} {\bibfnamefont {M.~N.}\ \bibnamefont
  {Barber}},\ }in\ \href@noop {} {\emph {\bibinfo {booktitle} {Phase
  Transitions and Critical Phenomena}}},\ Vol.~\bibinfo {Volume} {8},\ \bibinfo
  {editor} {edited by\ \bibinfo {editor} {\bibfnamefont {C.}~\bibnamefont
  {Domb}}\ and\ \bibinfo {editor} {\bibfnamefont {J.~L.}\ \bibnamefont
  {Lebowitz}}}\ (\bibinfo  {publisher} {Academic, London},\ \bibinfo {year}
  {1983})\ p.\ \bibinfo {pages} {145}\BibitemShut {NoStop}%
\bibitem [{\citenamefont {Privman}(1990)}]{privman:1990}%
  \BibitemOpen
  \bibfield  {author} {\bibinfo {author} {\bibfnamefont {V.}~\bibnamefont
  {Privman}},\ }in\ \href@noop {} {\emph {\bibinfo {booktitle} {Finite Size
  Scaling and Numerical Simulation of Statistical Systems}}},\ \bibinfo
  {editor} {edited by\ \bibinfo {editor} {\bibfnamefont {V.}~\bibnamefont
  {Privman}}}\ (\bibinfo  {publisher} {World Scientific, Singapore},\ \bibinfo
  {year} {1990})\ p.\ \bibinfo {pages {1}}\BibitemShut {NoStop}%
\bibitem [{\citenamefont {Pelissetto}\ and\ \citenamefont
  {Vicari}(2002)}]{pelissetto:2002}%
  \BibitemOpen
  \bibfield  {author} {\bibinfo {author} {\bibfnamefont {A.}~\bibnamefont
  {Pelissetto}}\ and\ \bibinfo {author} {\bibfnamefont {E.}~\bibnamefont
  {Vicari}},\ }\href {\doibase 10.1016/S0370-1573(02)00219-3} {\bibfield
  {journal} {\bibinfo  {journal} {Phys. Rep.}\ }\textbf {\bibinfo {Volume}
  {368}},\ \bibinfo {pages} {549} (\bibinfo {year} {2002})}\BibitemShut
  {NoStop}%
\bibitem [{\citenamefont {Krech}(1997)}]{Krech:1997}%
  \BibitemOpen
  \bibfield  {author} {\bibinfo {author} {\bibfnamefont {M.}~\bibnamefont
  {Krech}},\ }\href@noop {} {\bibfield  {journal} {\bibinfo  {journal} {Phys.
  Rev. E}\ }\textbf {\bibinfo {Volume} {56}},\ \bibinfo {pages} {1642}
  (\bibinfo {year} {1997})}\BibitemShut {NoStop}%
\bibitem [{\citenamefont {Krech}\ and\ \citenamefont
  {Dietrich}(1991)}]{Krech:1991}%
  \BibitemOpen
  \bibfield  {author} {\bibinfo {author} {\bibfnamefont {M.}~\bibnamefont
  {Krech}}\ and\ \bibinfo {author} {\bibfnamefont {S.}~\bibnamefont
  {Dietrich}},\ }\href@noop {} {\bibfield  {journal} {\bibinfo  {journal}
  {Phys. Rev. Lett.}\ }\textbf {\bibinfo {Volume} {66}},\ \bibinfo {pages}
  {345} (\bibinfo {year} {1991})}\BibitemShut {NoStop}%
\bibitem [{\citenamefont {Krech}\ and\ \citenamefont
  {Dietrich}(1992{\natexlab{a}})}]{Krech:1992a}%
  \BibitemOpen
  \bibfield  {author} {\bibinfo {author} {\bibfnamefont {M.}~\bibnamefont
  {Krech}}\ and\ \bibinfo {author} {\bibfnamefont {S.}~\bibnamefont
  {Dietrich}},\ }\href@noop {} {\bibfield  {journal} {\bibinfo  {journal}
  {Phys. Rev. A}\ }\textbf {\bibinfo {Volume} {46}},\ \bibinfo {pages} {1922}
  (\bibinfo {year} {1992}{\natexlab{a}})}\BibitemShut {NoStop}%
\bibitem [{\citenamefont {Krech}\ and\ \citenamefont
  {Dietrich}(1992{\natexlab{b}})}]{Krech:1992}%
  \BibitemOpen
  \bibfield  {author} {\bibinfo {author} {\bibfnamefont {M.}~\bibnamefont
  {Krech}}\ and\ \bibinfo {author} {\bibfnamefont {S.}~\bibnamefont
  {Dietrich}},\ }\href@noop {} {\bibfield  {journal} {\bibinfo  {journal}
  {Phys. Rev. A}\ }\textbf {\bibinfo {Volume} {46}},\ \bibinfo {pages} {1886}
  (\bibinfo {year} {1992}{\natexlab{b}})}\BibitemShut {NoStop}%
\bibitem [{\citenamefont {Evans}\ and\ \citenamefont
  {Stecki}(1994)}]{evans:1994}%
  \BibitemOpen
  \bibfield  {author} {\bibinfo {author} {\bibfnamefont {R.}~\bibnamefont
  {Evans}}\ and\ \bibinfo {author} {\bibfnamefont {J.}~\bibnamefont {Stecki}},\
  }\href {\doibase 10.1103/PhysRevB.49.8842} {\bibfield  {journal} {\bibinfo
  {journal} {Phys. Rev. B}\ }\textbf {\bibinfo {Volume} {49}},\ \bibinfo
  {pages} {8842} (\bibinfo {year} {1994})}\BibitemShut {NoStop}%
\bibitem [{\citenamefont {Borjan}\ and\ \citenamefont
  {Upton}(2008)}]{borjan:2008}%
  \BibitemOpen
  \bibfield  {author} {\bibinfo {author} {\bibfnamefont {Z.}~\bibnamefont
  {Borjan}}\ and\ \bibinfo {author} {\bibfnamefont {P.~J.}\ \bibnamefont
  {Upton}},\ }\href {\doibase 10.1103/PhysRevLett.101.125702} {\bibfield
  {journal} {\bibinfo  {journal} {Phys. Rev. Lett.}\ }\textbf {\bibinfo
  {Volume} {101}},\ \bibinfo {pages} {125702} (\bibinfo {year}
  {2008})}\BibitemShut {NoStop}%
\bibitem [{\citenamefont {Vasilyev}\ \emph {et~al.}(2007)\citenamefont
  {Vasilyev}, \citenamefont {Gambassi}, \citenamefont {Macio{\l}ek},\ and\
  \citenamefont {Dietrich}}]{Vasilyev:2007}%
  \BibitemOpen
  \bibfield  {author} {\bibinfo {author} {\bibfnamefont {O.}~\bibnamefont
  {Vasilyev}}, \bibinfo {author} {\bibfnamefont {A.}~\bibnamefont {Gambassi}},
  \bibinfo {author} {\bibfnamefont {A.}~\bibnamefont {Macio{\l}ek}}, \ and\
  \bibinfo {author} {\bibfnamefont {S.}~\bibnamefont {Dietrich}},\ }\href
  {\doibase 10.1209/0295-5075/80/60009} {\bibfield  {journal} {\bibinfo
  {journal} {EPL}\ }\textbf {\bibinfo {Volume} {80}},\ \bibinfo {pages} {60009}
  (\bibinfo {year} {2007})}\BibitemShut {NoStop}%
\bibitem [{\citenamefont {Vasilyev}\ \emph {et~al.}(2009)\citenamefont
  {Vasilyev}, \citenamefont {Gambassi}, \citenamefont {Macio{\l}ek},\ and\
  \citenamefont {Dietrich}}]{Vasilyev:2009}%
  \BibitemOpen
  \bibfield  {author} {\bibinfo {author} {\bibfnamefont {O.}~\bibnamefont
  {Vasilyev}}, \bibinfo {author} {\bibfnamefont {A.}~\bibnamefont {Gambassi}},
  \bibinfo {author} {\bibfnamefont {A.}~\bibnamefont {Macio{\l}ek}}, \ and\
  \bibinfo {author} {\bibfnamefont {S.}~\bibnamefont {Dietrich}},\ }\href
  {\doibase 10.1103/PhysRevE.79.041142} {\bibfield  {journal} {\bibinfo
  {journal} {Phys. Rev. E}\ }\textbf {\bibinfo {Volume} {79}},\ \bibinfo
  {pages} {041142} (\bibinfo {year} {2009})}{; \textit{ibid.}
  \textbf{80}, 039902(E) (2009)}\BibitemShut {NoStop}%
\bibitem [{\citenamefont {Hasenbusch}(2010{\natexlab{a}})}]{Hasenbusch:2010a}%
  \BibitemOpen
  \bibfield  {author} {\bibinfo {author} {\bibfnamefont {M.}~\bibnamefont
  {Hasenbusch}},\ }\href {\doibase 10.1103/PhysRevB.82.104425} {\bibfield
  {journal} {\bibinfo  {journal} {Phys. Rev. B}\ }\textbf {\bibinfo {Volume}
  {82}},\ \bibinfo {pages} {104425} (\bibinfo {year}
  {2010}{\natexlab{a}})}\BibitemShut {NoStop}%
\bibitem [{\citenamefont {Hasenbusch}(2010{\natexlab{b}})}]{Hasenbusch:2010}%
  \BibitemOpen
  \bibfield  {author} {\bibinfo {author} {\bibfnamefont {M.}~\bibnamefont
  {Hasenbusch}},\ }\href {\doibase 10.1103/PhysRevB.82.174434} {\bibfield
  {journal} {\bibinfo  {journal} {Phys. Rev. B}\ }\textbf {\bibinfo {Volume}
  {82}},\ \bibinfo {pages} {174434} (\bibinfo {year}
  {2010}{\natexlab{b}})}\BibitemShut {NoStop}%
\bibitem [{\citenamefont {Hasenbusch}(2010{\natexlab{c}})}]{Hasenbusch:2010b}%
  \BibitemOpen
  \bibfield  {author} {\bibinfo {author} {\bibfnamefont {M.}~\bibnamefont
  {Hasenbusch}},\ }\href {\doibase 10.1103/PhysRevB.82.174433} {\bibfield
  {journal} {\bibinfo  {journal} {Phys. Rev. B}\ }\textbf {\bibinfo {Volume}
  {82}},\ \bibinfo {pages} {174433} (\bibinfo {year}
  {2010}{\natexlab{c}})}\BibitemShut {NoStop}%
\bibitem [{\citenamefont {Diehl}(1997)}]{Diehl:1997}%
  \BibitemOpen
  \bibfield  {author} {\bibinfo {author} {\bibfnamefont {H.~W.}\ \bibnamefont
  {Diehl}},\ }\href@noop {} {\bibfield  {journal} {\bibinfo  {journal} {Int. J.
  Mod. Phys. B}\ }\textbf {\bibinfo {Volume} {11}},\ \bibinfo {pages} {3503}
  (\bibinfo {year} {1997})}\BibitemShut {NoStop}%
\bibitem [{\citenamefont {Binder}\ and\ \citenamefont
  {Hohenberg}(1972)}]{Binder:1972}%
  \BibitemOpen
  \bibfield  {author} {\bibinfo {author} {\bibfnamefont {K.}~\bibnamefont
  {Binder}}\ and\ \bibinfo {author} {\bibfnamefont {P.~C.}\ \bibnamefont
  {Hohenberg}},\ }\href@noop {} {\bibfield  {journal} {\bibinfo  {journal}
  {Phys. Rev. B}\ }\textbf {\bibinfo {Volume} {6}},\ \bibinfo {pages} {3461}
  (\bibinfo {year} {1972})}\BibitemShut {NoStop}%
\bibitem [{\citenamefont {Diehl}\ and\ \citenamefont
  {Smock}(1993)}]{Diehl:1993}%
  \BibitemOpen
  \bibfield  {author} {\bibinfo {author} {\bibfnamefont {H.~W.}\ \bibnamefont
  {Diehl}}\ and\ \bibinfo {author} {\bibfnamefont {M.}~\bibnamefont {Smock}},\
  }\href {\doibase 10.1103/PhysRevB.47.5841} {\bibfield  {journal} {\bibinfo
  {journal} {Phys. Rev. B}\ }\textbf {\bibinfo {Volume} {47}},\ \bibinfo
  {pages} {5841} (\bibinfo {year} {1993})}{; \textit{ibid.}
  6740 (1993)}\BibitemShut {NoStop}%
\bibitem [{\citenamefont {Zinn-Justin}(2002)}]{zinn-justin:2002}%
  \BibitemOpen
  \bibfield  {author} {\bibinfo {author} {\bibfnamefont {J.}~\bibnamefont
  {Zinn-Justin}},\ }\href@noop {} {\emph {\bibinfo {title} {Quantum Field
  Theory and Critical Phenomena}}},\ \bibinfo {edition} {4th}\ ed.\ (\bibinfo
  {publisher} {Clarendon Press, Oxford},\ \bibinfo {year} {2002})\BibitemShut
  {NoStop}%
\bibitem [{\citenamefont {Kondrat}\ \emph {et~al.}(2009)\citenamefont
  {Kondrat}, \citenamefont {Harnau},\ and\ \citenamefont
  {Dietrich}}]{Kondrat:2009}%
  \BibitemOpen
  \bibfield  {author} {\bibinfo {author} {\bibfnamefont {S.}~\bibnamefont
  {Kondrat}}, \bibinfo {author} {\bibfnamefont {L.}~\bibnamefont {Harnau}}, \
  and\ \bibinfo {author} {\bibfnamefont {S.}~\bibnamefont {Dietrich}},\ }\href
  {\doibase 10.1063/1.3259188} {\bibfield  {journal} {\bibinfo  {journal} {J.
  Chem. Phys.}\ }\textbf {\bibinfo {Volume} {131}},\ \bibinfo {eid} {204902}
  (\bibinfo {year} {2009})}\BibitemShut {NoStop}%
\bibitem [{\citenamefont {Hanke}\ and\ \citenamefont
  {Dietrich}(1999)}]{Hanke:1999a}%
  \BibitemOpen
  \bibfield  {author} {\bibinfo {author} {\bibfnamefont {A.}~\bibnamefont
  {Hanke}}\ and\ \bibinfo {author} {\bibfnamefont {S.}~\bibnamefont
  {Dietrich}},\ }\href {\doibase 10.1103/PhysRevE.59.5081} {\bibfield
  {journal} {\bibinfo  {journal} {Phys. Rev. E}\ }\textbf {\bibinfo {Volume}
  {59}},\ \bibinfo {pages} {5081} (\bibinfo {year} {1999})}\BibitemShut
  {NoStop}%
\bibitem [{\citenamefont {Harnau}\ \emph {et~al.}(2004)\citenamefont {Harnau},
  \citenamefont {Penna},\ and\ \citenamefont {Dietrich}}]{harnau:2004}%
  \BibitemOpen
  \bibfield  {author} {\bibinfo {author} {\bibfnamefont {L.}~\bibnamefont
  {Harnau}}, \bibinfo {author} {\bibfnamefont {F.}~\bibnamefont {Penna}}, \
  and\ \bibinfo {author} {\bibfnamefont {S.}~\bibnamefont {Dietrich}},\
  }\href@noop {} {\bibfield  {journal} {\bibinfo  {journal} {Phys. Rev. E}\
  }\textbf {\bibinfo {Volume} {70}},\ \bibinfo {pages} {021505} (\bibinfo
  {year} {2004})}\BibitemShut {NoStop}%
\end{thebibliography}
\end{document}